 \documentclass[5p,twocolumn,sort&compress]{elsarticle}

 \usepackage{natbib}
 \usepackage[colorlinks=true,linkcolor=black, citecolor=blue, urlcolor=blue]{hyperref}
 \usepackage{ucs}
 \usepackage[utf8x]{inputenc}
 \usepackage{latexsym}
 \usepackage{amsfonts}
 \usepackage{amsmath}
 \usepackage{color}
 \usepackage{bm}
 \usepackage[normalem]{ulem}

 \def\bH{{\bf H}}

 \title{Coherent dynamics in cavity femtochemistry:
       application of the multi-configuration
       time-dependent Hartree method.}

 \date{\today}

 \author[ov]{Oriol Vendrell\corref{cor1}}
 \ead{oriol.vendrell@phys.au.dk}

 \cortext[cor1]{Corresponding author\\
 \textit{Email address:}
 \texttt{\url{oriol.vendrell@phys.au.dk}}}

 \address[ov]{Department of Physics and Astronomy,
              Aarhus University,
              Ny Munkegade 120, 8000 Aarhus C, Denmark}

\begin{document}

 \begin{keyword}
    cavity electrodynamics
    \sep
    femtochemistry
    \sep
    superradiance
    \sep
    quantum dynamics
 \end{keyword}

 \bibliographystyle{elsarticle-num}


 \begin{abstract}
The photochemistry of a molecular ensemble coupled to a resonance cavity and
triggered by a femtosecond laser pulse is investigated from a real-time,
quantum dynamics perspective with the multi-configuration time-dependent
Hartree method.  Coherent excitation of a superposition of electronic states in
the ensemble leads to superradiant energy transfer to the cavity
characterized by quadratic scaling with the number of molecules. Electronic
decoherence associated with loss of nuclear wave packet overlap among those
states destroys superradiant energy transfer, returning to a
linear regime. For equal pump laser conditions, the photoexcitation probability
per molecule decreases with increase of the number of molecules inside the cavity.
This is caused by a loss of resonance condition of the laser with the
bright electronic-photonic states of the coupled cavity-ensemble system.
Increase of the laser bandwidth restores the energy transferred per molecule
and the trigger probability remains independent of the number of molecules
in the cavity.
 \end{abstract}


 \maketitle

\section{Introduction} \label{sec:introduction}

%
The interaction of atoms~\cite{jay63:89,tav67:714,har89:24,mil05:551,wal06:1325}
and ions~\cite{her09:494} with quantized light has long been well known and
investigated in depth.
In recent years, interest for the fundamental properties and technological
applications of hybrid light-matter systems of molecular nature is quickly
raising, which is motivated to a large extent by the high tunability and
relative ease of preparation of such systems
~\cite{ke08:116401,sch11:196405,sch13:125,hut12:1624,geo15:281,ebb16:2403}.
From a \textit{chemical} perspective, hybrid systems composed of a molecular
ensemble coupled to a resonance cavity can lead to novel strategies to
steer~\cite{mor07:73001,sch11:196405,hut12:1624,gal15:41022,her16:238301,ebb16:2403,%
kow16:2050,gal16:13841} and spectroscopically probe~\cite{car12:125424} the
molecular properties and response of their individual members by exploiting
their collective coupling to the cavity.

%
Recent theoretical investigations have elucidated cavity effects on the
non-adiabatic molecular dynamics of a single coupled
molecule~\cite{kow16:2050,kow16:54309}.
The effect on bonding and electronic structure parameters of molecules in a
cavity has been as well the subject of recent
investigations~\cite{tok13:233001,rug14:12508,gal15:41022,cwi16:33840,fli17:3026}.
As expected, such investigations confirm that structural properties such as bond
length or orientation are, to a large extent, related to the individual coupling
strength of each molecule to the
cavity.

Specially interesting are theoretical proposals to exploit collective coupling
effects as a way to alter the chemical evolution of the molecules in the cavity. These
may involve either quenching~\cite{gal16:13841} or enhancing~\cite{gal17:136001}
photochemical reactions in excited electronic states, in which the presence of a
molecular ensemble becomes important.
%
%
In Ref.~\cite{gal16:13841}, for example, it was proposed that a photochemical
reaction can be suppressed by the admixture of ground state character, which
features a potential energy barrier, to some of the excited polaritonic (i.e.
involving coupled electronic and photonic degrees of freedom) potential energy
surfaces (PES), whereby an initial wavepacket was assumed to evolve on the
lowest polaritonic PES after an instantaneous trigger process.
In this respect, there is experimental evidence that, at least for reaction
rates, the alteration of chemically relevant properties by a cavity is
possible~\cite{hut12:1624}.

The quick increase in dimensionality of the Hilbert space
when describing molecular-cavity problems involving a molecular ensamble
has lead to the application of different kinds of theory approaches.
On the one hand,
an adiabatic separation of nuclear and electronic and photonic degrees of
freedom can be invoked, which leads to the construction of
polaritonic PES. To obtain those, the polaritonic
Hamiltonian parametrized by the nuclear positions
is diagonalized, very often in the single molecular electronic excitation
space, such that the
rank of the Hamiltonian matrix grows linearly
with the number of molecules
$N$~\cite{gal16:13841,luk17:4324,gal17:136001}.
Therefore, such treatments have been able to account for situations with a low
excitation density -- i.e. a low number of cavity photons per molecule.

%
On the other hand, full quantum dynamics studies based on the standard method, meaning a
product-grid representation of the wavefunction for all degrees of freedom,
have been reported~\cite{gal15:41022,kow16:2050,kow16:54309,fli17:3026}.
These approaches naturally account
for quantum evolution of nuclear, electronic and photonic
degrees of freedom. They face, however, a dramatic exponential scaling
when treating $N$-molecule ensembles in a cavity and therefore
have been limited to the
description of one~\cite{kow16:2050,kow16:54309,fli17:3026}
or two molecules~\cite{gal15:41022} coupled to the electromagnetic modes.

The description of hybrid systems in terms of polaritonic PES has been
recently extended to a surface-hopping algorithm for classical nuclear
trajectories~\cite{tul90:1061} to account for polaritonic transitions~\cite{luk17:4324}.
This opens interesting possibilities to follow the time evolution and relaxation
of polaritons in a cavity for large molecular numbers, and for
complex molecules.
Nonetheless, theoretical descriptions based on classical mechanics
for the nuclei are not able to account for the coherent quantum
evolution among members of the molecular ensemble as prepared, e.g., by
femtosecond pump lasers, and offer a limited description of electronic decoherence
processes~\cite{Sub11:19}, which may be important at short times upon
photo-excitation~\cite{arn17:33425,vac17:83001}.

For example, a challenging scenario is presented by the
coherent excitation of \textit{all} members of a molecular ensemble
by a femtosecond laser pulse to produce, for each of the molecules, a
superposition of two (or more) electronic states.
This situation leaves the ensemble in a ``cooperative" state,
as termed by Dicke in his seminal work~\cite{dic54:99},
which can lead to superradiant
energy transfer to electromagnetic modes.
States of this kind have been experimentally achieved recently for an ensemble
of as many as $10^{13}$ spins in a resonance cavity, which were shown to couple to
the cavity mode in the superradiant regime~\cite{ros17:31002}.
The theoretical consideration of analogous situations with molecules
requires the inclusion of the nuclear degrees of freedom besides
accounting for the full space of polaritonic states with
2$^N$ molecular excitations, for $N$ molecules and two electronic states per
molecule.


In this work, full wavepacket quantum dynamics of an $N$-molecule ensemble
coupled to a cavity and ranging from low to high
excitation densities is considered and efficiently
solved.
Thereby, collective responses follow exclusively from the
    coupled
dynamics of the
separate constituents
    and no
collective parameters, e.g., the Rabi splitting of the whole ensemble,
enter the Hamiltonian.
The theoretical description is free from the construction of
polaritonic adiabatic PES (and their couplings)
on which the nuclei evolve. Instead nuclear, electronic and photonic degrees
of freedom are propagated quantum dynamically on the same footing for all
members of the ensemble and the cavity.
In this way,
the models investigated explicitly consider
the coupling of each molecule
to the cavity and to external fields, e.g., femtosecond laser pulses pumping or
driving the system.

This level of description is achieved by employing the highly efficient
multi-configuration
time-dependent Hartree (MCTDH) method for the propagation of multidimensional
wavepackets~\cite{mey90:73,man92:3199,bec00:1,mey03:251}.
As it will be discussed,
the tensorial contraction of the wave function in the primitive grid basis
representation inherent to MCTDH (and its multilayer extension)
ideally suits the description of a molecular ensemble coupled to a cavity,
thus accounting for quantum correlations among the ensemble members and, if
needed, for large photon numbers in the cavity.

This theoretical framework will be used to shed light on fundamental aspects of
the interaction of the molecular ensemble with the cavity.
In Sec.~\ref{sec:laser} the
short time dynamics of the molecular ensemble under coherent excitation by femtosecond
laser pulses will be described in detail.
Section~\ref{sec:coherence} analyzes the mechanism of coherent energy transfer
between the ensemble and the cavity, and the role of electronic coherence,
whereas Sec.~\ref{sec:photons} explores situations in which the number of
photons in the cavity is similar or larger than the number of molecules in the
ensemble.
Before this, Sec.~\ref{sec:theory} lays down the theoretical framework
for the description of ensemble-cavity systems
and for the application of the MCTDH method to compute their quantum dynamical
evolution.

\section{Theoretical Framework and Computational Details}
    \label{sec:theory}

    \subsection{Molecular ensemble-cavity Hamiltonian}

    Our subject of investigation is an ensemble of molecules
    placed inside a cavity that supports quantized
    electromagnetic modes. The molecular density is assumed low enough that
    molecule-molecule interactions can be
    neglected~\cite{dic54:99,jay63:89,tav67:714}. Hence each molecule is coupled
    solely to the quantized electromagnetic modes of the cavity
    and to external electromagnetic radiation, -- e.g. a laser pulse --
    which is described classically.

    The basic form of the Hamiltonian for such scenario is well known and
    has been presented
    elsewhere~\cite{gal15:41022,kow16:54309,kow16:2050,fli17:3026}.
    Here, the development of Ref.~\cite{fli17:3026} is closely followed.
    Compared to other treatments, it keeps
    a quadratic dipole self-energy term~\cite{Faisal_1987} in the light-matter
    interaction (see below), which becomes only relevant in the ultra-strong coupling
    limit. In our case we keep this term for completeness but it has no effect
    on the dynamics for the range of conditions investigated.

    Usually
    a Rabi-type term~\cite{rab37:652} describes the coupling of each
    molecule to the electromagnetic
    modes via a nuclear-position dependent dipole
    ~\cite{gal15:41022,kow16:2050,fli17:3026}. The Jaynes-Cummings~\cite{jay63:89}
    (and Tavis-Cummings for an ensemble~\cite{tav67:714})
    Hamiltonian follows if one adopts the
    rotating wave approximation, employed e.g., in
    Refs.~\cite{her16:238301,kow16:54309,luk17:4324}, but not in this work.

    In the following, the Hamiltonian of the ensemble-cavity system is introduced
    with emphasis on the aspects relevant to this work.
    The Hamiltonian for the hybrid system is given by
    \begin{align}
        \label{eq:htot}
        \hat{H}  = \sum_{l=1}^{N} \hat{H}^{(l)}_{\mathsf{mol}}
                  + \hat{H}_{\mathsf{cav}}
                  + \hat{H}_{\mathsf{las}},
    \end{align}
    where the Hamiltonian of the $l$-th molecule
    \begin{align}
        \label{eq:hmol1}
        \hat{H}^{(l)}_{\mathsf{mol}} & =  \hat{T}^{(l)}_n + \hat{H}^{(l)}_e
    \end{align}
    is written as a sum of its nuclear kinetic energy $\hat{T}^{(l)}_n$ and
    its clamped nuclei Hamiltonian $\hat{H}^{(l)}_e$,
    including the electronic kinetic energy and all intra-molecular Coulombic
    interactions~\cite{wor04:127}.
    The Hamiltonian of the quantized electromagnetic modes
    and their interaction with the molecules,
    \begin{align}
        \label{eq:hcav}
        \hat{H}_{\mathsf{cav}} =
             \frac{1}{2}\left[
                 \hat{p}^2 +
                 \omega_c^2 \left(
                    \hat{q} + \frac{\vec{\lambda}\cdot\hat{\vec{D}}}{\omega_c}
                 \right)^2
             \right],
    \end{align}
    is given in the harmonic oscillator
    form, in length gauge and in
    dipole approximation~\cite{kow16:2050,fli17:3026}.
    Only one mode is considered for the sake of clarity.
    $\omega_c$ is the angular frequency of the cavity
    mode and $\vec{\lambda}$ is the dipole coupling strength
    $\lambda=\sqrt{1/\epsilon_0 V}$
    times the unit
    polarization vector in the given
    cavity. $\hat{\vec{D}}$ is the dipole operator acting on all matter
    degrees of freedom and $V$ is the volume of the quantized
    cavity mode.
    In Eq.~(\ref{eq:hcav}) one can substitute
    $\hat{q}=\sqrt{\hbar/2\omega_c}(\hat{a}^\dagger + \hat{a})$
    and
    $\hat{p}=i\sqrt{\hbar\omega_c/2}(\hat{a}^\dagger - \hat{a})$
    and expand the quadratic term to reach~\cite{Faisal_1987}
    \begin{align}
        \label{eq:hcav2}
        \hat{H}_{\mathsf{cav}}
             & =
             \hbar\omega_c \left( \frac{1}{2} + \hat{a}^\dagger\hat{a} \right)
             +
             \sqrt{\frac{\hbar\omega_c}{2}} \vec{\lambda}\cdot\hat{\vec{D}}
             \left( \hat{a}^\dagger + \hat{a} \right) \\\nonumber
             & +
             \frac{1}{2} \left( \vec{\lambda}\cdot\hat{\vec{D}} \right)^2
    \end{align}
    where $\hat{a}^\dagger$ and $\hat{a}$ are the photon creation and
    annihilation operators, respectively.
    The interaction with external fields, e.g., a femtosecond laser pulse, is
    treated semiclassically
    \begin{align}
        \label{eq:hlas}
        \hat{H}_{\mathsf{las}} = \vec{E}(t) \hat{\vec{D}},
    \end{align}
    in the dipole approximation and in length gauge, where
    $\vec{E}(t)$ is the time dependent
    electric field of the laser.
    The total dipole is simply the sum of the individual molecular
    dipoles
    \begin{align}
        \label{eq:dip}
        \hat{\vec{D}} = \sum_{l=1}^{N} \hat{\vec{D}}^{(l)},
    \end{align}
    which in terms of the nuclear (uppercase) and electronic (lowercase)
    coordinates of each molecule read
    \begin{align}
        \label{eq:dipmol}
        \hat{\vec{D}}^{(l)} = q_e \left( \sum_{\alpha=1}^{N_n}
        Z_\alpha\vec{R}_\alpha^{(l)} - \sum_{\beta=1}^{N_e}  \vec{r}_\beta^{(l)}
                              \right)
    \end{align}
    with $q_e$ the magnitude of the electron charge.

    At this point, and without loss of generality, electronic adiabatic
    eigenstates
    $|\psi_s^{(l)}(\mathbf{R}^{(l)})\rangle$
    of the molecular clamped nuclei Hamiltonians
    $\hat{H}^{(l)}_e$ are introduced
    \begin{align}
        \label{eq:tisemol}
        \hat{H}^{(l)}_e
        |\psi_s^{(l)}(\mathbf{R}^{(l)})\rangle
        =
        V_s^{(l)}(\mathbf{R}^{(l)})
        |\psi_s^{(l)}(\mathbf{R}^{(l)})\rangle.
    \end{align}
    In the following, their parametric
    dependence on the nuclear coordinates $(\mathbf{R}^{(l)})$ is dropped for
    the sake of clarity.
    %
    Projection of the $l$-th molecular Hamiltonian~(\ref{eq:hmol1})
    onto the corresponding electronic basis
    \begin{align}
        \label{eq:hmol2prj}
        \hat{H}^{(l)}_{\mathsf{mol}} \equiv
        \sum_{s=1}^{N_s} \sum_{r=1}^{N_s}
        |\psi_s^{(l)}\rangle\langle\psi_s^{(l)}|
        \hat{H}^{(l)}_{\mathsf{mol}}
        |\psi_r^{(l)}\rangle\langle\psi_r^{(l)}|
    \end{align}
    introduces the matrix elements of the nuclear Hamiltonian
    \begin{align}
        \label{eq:hmol2bo}
        \left[ \hat{H}^{(l)}_{\mathsf{mol}} \right]_{sr}
        & =
        \langle \psi_s^{(l)}| \hat{H}^{(l)}_{\mathsf{mol}} |\psi_r^{(l)}\rangle \\\nonumber
        &=
        \hat{T}^{(l)}_n + \hat{\Lambda}_{sr}^{(l)}
        + \hat{V}_s^{(l)}\delta_{sr},
    \end{align}
    in terms of adiabatic PES $\hat{V}_s^{(l)}$
    and their non-adiababtic
    couplings
    $\hat{\Lambda}_{sr}^{(l)} =
    \langle\psi_s^{(l)}|\hat{T}_n^{(l)}|\psi_r^{(l)}\rangle -
    \hat{T}_n^{(l)}\delta_{sr}$~\cite{wor04:127}.
    The brackets above denote integration over the electronic coordinates only.
    %
    %
    Similarly, projection
    of the dipole operator
    onto the same electronic basis yields
    \begin{align}
        \label{eq:dipbo1}
             \left[ \hat{\vec{D}} \right]_{sr}^{(l)} =
                     \langle \psi_s^{(l)}| \hat{\vec{D}} |\psi_r^{(l)}\rangle
             = \hat{\vec{\mu}}_{sr}^{(l)}(\mathbf{R}^{(l)}),
    \end{align}
    where $\hat{\vec{\mu}}_{sr}^{(l)}$ is the position dependent dipole (or
    transition dipole for $s\neq r$) matrix element of the $l$-th molecule,
    obtainable, typically, from quantum chemistry calculations.
    No
    intermolecular
    dipole terms are considered in Eq.~(\ref{eq:dipbo1})
    because of the assumed zero overlap among electronic states of different molecules.
    (See discussion at the beginning of this section).

    Throughout this work the molecules will be assumed to be aligned
    with the polarization axis of the cavity mode and of the external laser
    field, and the vector notation for the electric
    field and dipole operator will be dropped accordingly.
    The various parameters of the model that are relevant to each
    part of this investigation will be introduced as needed.

    \subsection{Quantum dynamics with MCTDH}
    \label{sec:mctdh}

    The MCTDH method is an efficient approach to propagate the time-dependent
    Schrödinger equation for multi-dimensional systems. The method was first
    introduced to treat the multi-dimensional quantum dynamics of molecular
    systems~\cite{mey90:73,man92:9062}.
    Over the years, though, applications to other types of problems, often
    linked to further developments of the basic theory, were successfully pushed
    forward, for example to describe the dynamics of
    electrons~\cite{cai05:12712,hax11:63416,miy14:63416} and bosonic
    systems~\cite{alo07:154103,alo08:33613}, to name just a few.  An in-depth
    review of the basic theory can be found in Ref.~\cite{bec00:1}.
    The quantum dynamics calculations in this work have been performed with the
    Heidelberg MCTDH package version v$85.5$~\cite{mctdh:MLpackage}.

    The basic theory is briefly described in this Section, where
    the usual nomenclature in the MCTDH literature is used
    for consistency~\cite{bec00:1}.
    The MCTDH ansatzt specialized to the ensemble-cavity case is discussed
    in Sec.~\ref{sec:wfn}.
    The MCTDH ansatzt for the wave function reads
    \begin{eqnarray}
    \label{eq2:ansatz1}
        \nonumber
        |\Psi(q_1,\dots,q_f,t)\rangle & = &
        \sum_{j_1}^{n_1}\ldots\sum_{j_f}^{n_f}{A_{j_1\ldots j_f}(t)
        \prod_{\kappa=1}^{f}{|\varphi_{j_{\kappa}}^{(\kappa)}(q_{\kappa},t)\rangle}} \\
        & = &  \sum_{J}{A_J(t)|\Phi_J(t)\rangle}\,,
    \end{eqnarray}
    where $A_J(t)$ is the time-dependent expansion coefficient
    of the $J$-th configuration labeled with multi-index $J$, and
    $|\Phi_J(t)\rangle$ is the $J$-th time-dependent Hartree product, which is a
    direct product of single-particle functions (SPFs) for each degree of
    freedom. These are analogous to molecular orbitals in electronic structure
    theory~\cite{szabo-book}.
    After applying a time-dependent variational principle to this ansatzt the
    MCTDH equations of motion
    \begin{eqnarray}
    \label{eq:eom}
    \mathrm{i}\dot{A}_J &=& \sum_{L}\langle\Phi_J\left|H\right|\Phi_L\rangle A_L\,,\\
    \mathrm{i}\dot{{\bf \varphi}}^{(\kappa)}
         &=& (1-P^{(\kappa)})({\bf\rho}^{(\kappa)})^{-1}\langle\bH\rangle^{(\kappa)}
         {\bf \varphi}^{(\kappa)}\,\nonumber
    \end{eqnarray}
    are obtained.
    Here a vector notation
    ${\bf  \varphi}^{(\kappa)}=(|\varphi_1^{(\kappa)}\rangle,\ldots
    |\varphi_{n_{\kappa}}^{(\kappa)}\rangle)^T$
    is used,
    \begin{equation}
      \label{eq:projector}
      P^{(\kappa)}=
     \sum_{j=1}^{n_{\kappa}}|{\bf \varphi}^{(\kappa)}_j\rangle\langle{\bf \varphi}^{(\kappa)}_j|\,
    \end{equation}
    is the projector on the space spanned by the SPFs for the $\kappa$th degree
    of freedom, and $\langle \bH\rangle^{(\kappa)}$ and $\rho^{(\kappa)}$ are
    mean-fields and the density matrix~\cite{bec00:1}.

    The SPFs are expanded in turn in a time-independent basis for each degree of
    freedom
    \begin{align}
      \label{eq:spf}
     |{\varphi}^{(\kappa)}_j\rangle =
        \sum_{i=1}^{N_\kappa} c_{i,j}^{(\kappa)}(t) |\chi_i^{(\kappa)}\rangle,
    \end{align}
    where, for convenience, very often the states of the primitive representation
    $|\chi_i^{(\kappa)}\rangle$ are taken from a
    discrete variable representation~\cite{bec00:1}.

    The efficiency gain in MCTDH compared to propagating directly in the
    primitive basis (the standard method) arises from the usually big
    difference between the size of the primitive space
    $\prod_{\kappa=1}^{f}N_\kappa$, and the size of the configuration space
    $\prod_{\kappa=1}^{f}n_\kappa$ needed to achieve convergence
    in the correlated dynamics.
    Mode combination~\cite{mey03:251} (used
    here) and especially multilayer
    MCTDH~\cite{wan03:1289,man08:164116,Ven11:44135,cao13:134103}, which is implemented in the
    Heidelberg MCTDH package~\cite{Ven11:44135},
    can boost even further the efficiency of
    the method allowing, in favorable cases, the description of hundreds to
    thousands of degrees of freedom~\cite{cra07:144503,wan08:139,Ven11:44135}.

    \subsection{MCTDH for the ensemble-cavity problem}
    \label{sec:wfn}

\begin{figure}[t!]
       \begin{center}
          \includegraphics[width=0.95\columnwidth]{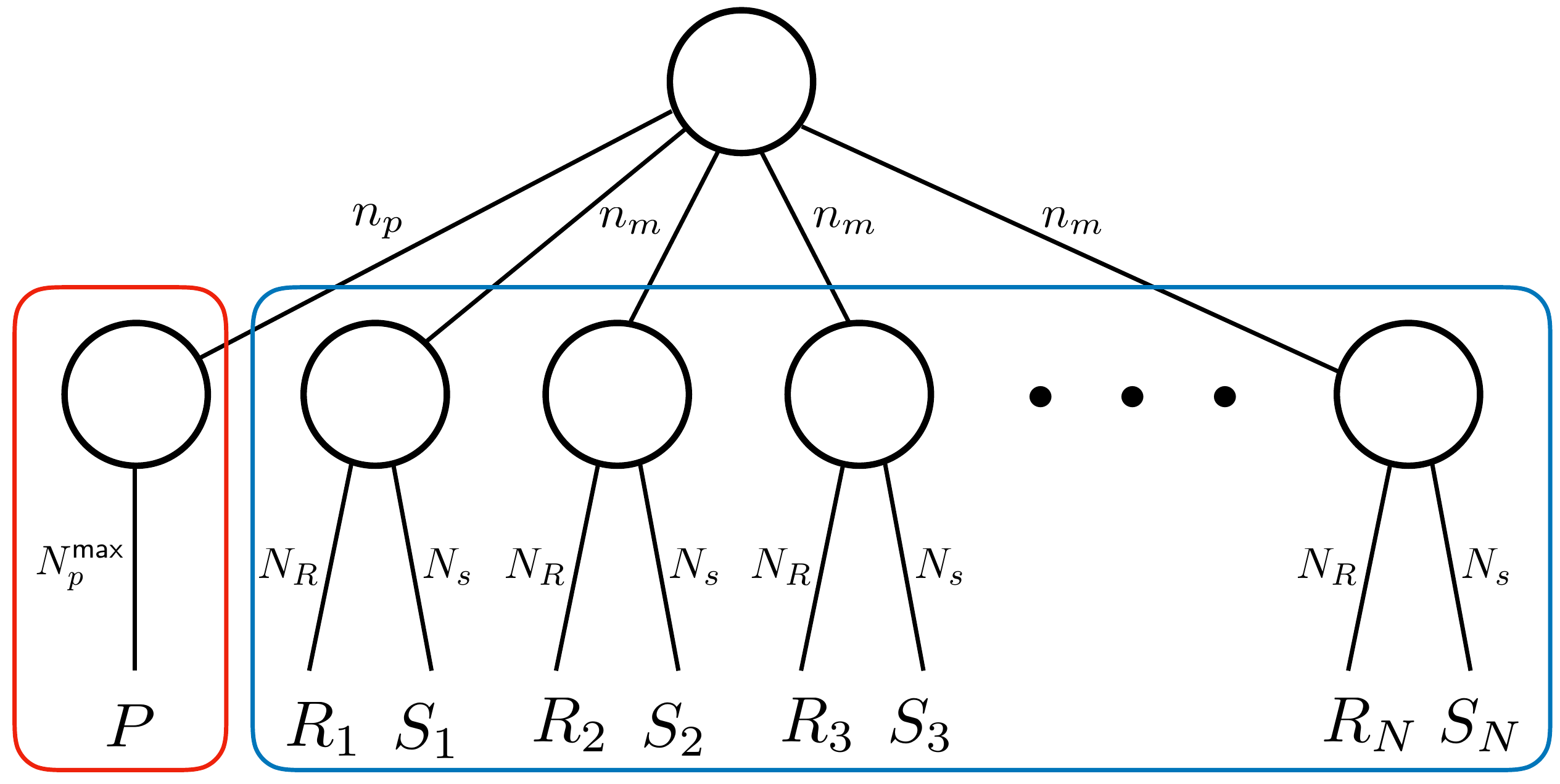}
       \end{center}
       \caption{Tree representation of the MCTDH wave functions
        used to represent the cavity mode and the molecular ensemble.
        The red box marks the photonic mode and the blue box
        marks the molecular degrees of freedom.
        $N_R$ and $N_s$ refer to the number of primitive basis functions (grid
        points in this particular case) and number of electronic states in each
        molecule, respectively.
        The number of single particle functions for the molecules and for the
        cavity mode are $n_m$ and $n_p$, respectively.
        $P$, $R_l$ and $S_l$ denote the cavity, nuclear and electronic degrees
        of freedom, respectively.
        }
       \label{fig:wftree}
    \end{figure}

    Upon projection of the molecular Hamiltonians onto an electronic basis for
    each molecular system in (\ref{eq:hmol2prj}), the total MCTDH wave function
    for the ensemble-cavity system molecules takes the form
    \begin{align}
        \label{eq:wfn0}
        |\Psi(t)\rangle &= \sum_{j_1,\ldots,j_N,j_p}^{n_1,\ldots,n_N,n_p}
        A_{j_1,\ldots,j_N,j_p}(t) \cdot \\\nonumber
        & \prod_{l=1}^N
        \Bigg(
            \sum_{s_l=1}^{N_s} \phi_{s_l,j_l}^{(l)}(t)|\psi_{s_l}^{(l)}\rangle
        \Bigg)
        \Bigg(
            \sum_{P=1}^{N_p} B_{P,j_p}(t) |P\rangle
        \Bigg).
    \end{align}
    Here the $n_l$ and $n_p$ are the number of SPF basis for each molecule
    and for the cavity mode, respectively, where nuclear and electronic degrees
    of freedom are combined to one logical mode (cf. Fig.~\ref{fig:wftree}).
    As before, $N_s$ is the number of
    relevant electronic states per molecule and $N_p$
    is the maximum number of photons allowed in the cavity.
    The $\phi_{s_l,j_l}^{(l)}(t)$ functions are the nuclear wave packets
    for molecule $l$ in electronic state $s_l$, whereby index $j_l$ refers to
    the configuration space as specified by $A_{j_1,\ldots,j_N,j_p}$ in
    Eq.~(\ref{eq:wfn0}).
    $B_{P,j_c}(t)$ are the expansion coefficients of the primitive photonic
    space for $P$ cavity photons with configuration space index $j_p$.
    In absence of coupling between the molecules and the cavity, and of direct
    couplings between the molecules, one single Hartree product suffices in
    Eq.~(\ref{eq:wfn0}) and all $\{n_l,n_p\}$ become equal to one.

    Figure~\ref{fig:wftree} presents a tree representation of the MCTDH
    wave function. Such representations become particularly useful to describe
    multilayer wave functions, but are also very illustrative in the normal (two
    layer) MCTDH case.
    The top node represents the $A_{j_1,\ldots,j_N,j_p}(t)$ tensor, with one line
    per index. The nodes at the bottom layer represent the
    expansion coefficients of the molecular functions
    $\phi_{s_l,j_l}^{(l)}(t)$ and the $B_{P,j_p}(t)$ coefficients of the
    cavity mode.
    The left-most box in Fig.~\ref{fig:wftree} encloses the photonic degree of
    freedom. Since the quantized mode of the cavity is represented in its
    harmonic oscillator form, naturally a
    DVR constructed from harmonic oscillator eigenfunctions~\cite{bec00:1}
    is used to represent the photonic states.
    The size of the primitive basis $N_p$ determines the maximum
    number of photons in the cavity mode and has been set to values between 10
    and 60, depending on the details of the corresponding calculation. A
    substantially larger number of cavity photons or further cavity modes
    could be accommodated with ease if required.
    %

    %
    The right-most box in Fig.~\ref{fig:wftree} encloses the molecular degrees
    of freedom.
    In this application we focus on diatomic molecules with two electronic
    states each.
    The interatomic distance $R_l$ is
    discretized using a Fourier basis, and the electronic degree of freedom
    $S_l$
    is truly discrete
    with possible values, in this particular example (see below), $0$ and $1$.
    As already mentioned, the two degrees of freedom are here combined into one logical mode per
    molecule. The $R$ grids have 4096 points each, which are necessary to
    describe the dissociative dynamics of the NaI system (introduced below)
    and to represent the
    momentum achieved by the relatively heavy atoms. Therefore each molecule's
    primitive representation consists of 8192 grid points.
    The number of SPFs per mode required for converged results varies, but
    $n_\kappa$ about 5 provides converged results in most of the
    wave packet propagations presented below.
    %
    Finally, in MCTDH jargon,
    each molecule is described in a single-set formulation~\cite{mey03:251},
    which is the only practical alternative when each molecule carries its own
    electronic state index.

    In terms of computational effort and scalability of the approach, technical
    details of the largest calculations in this study are reported in
    Table~\ref{tab:mctdh} and correspond to either 5 or 8 molecular systems and
    one cavity mode, referred for brevity as calculations C$_5$ and C$_8$,
    respectively.
    C$_5$ is dominated by the number of SPF coefficients, where most of the
    propagation time is spent, whereas the array of $A_J$ coefficients is still
    short relative to typical MCTDH applications.
    This is due to the large size of the primitive grid needed to describe the
    photodissociation of NaI.
    C$_8$ is a more balanced case, with an $A_J$ array of order $10^5$ entries
    and almost equal propagation time spent in the coefficients and the SPFs. In
    usual applications, order $10^6$ $A_J$ coefficients are manageable, whereas
    order $10^7$ becomes hard to propagate. This means, for this particular
    investigation, that 9 to 10 molecules can be tretaed at a good level of
    accuracy.
    Clearly, standard method calculations on the corresponding primitive grids
    are impossible in the foreseeable future.

    As already mentioned, the calculations reported here are based
    on the normal (2-layer) MCTDH approach.
    Going beyond this system size, both in terms of molecular
    complexity and molecular number, requires using
    the multilayer MCTDH
    algorithm~\cite{wan03:1289,man08:164116,Ven11:44135,cao13:134103}.
    The Heidelberg implementation of multilayer MCTDH~\cite{mctdh:MLpackage}
    has been recently applied to describe the photophysics of
    naphtalene (48 D) and anthracene (66 D) molecules in full
    dimensionality~\cite{men13:14313}, as well as
    models of light-harvesting-complexes with hundreds of
    modes~\cite{shi17:184001}. This indicates that the extension to even larger
    ensembles of high-dimensional molecular systems than will be discussed here should
    be within reach and will be the subject of future work.

    \begin{table}[!]
        \begin{tabular}{cccccc}
            \hline
         &   $N_\Psi$
         & $N_A$
         & $N_\textrm{prim}$
         & $t_\textrm{CPU}$/$N_\textrm{fs}$[s] \\
            \hline
           C$_5$
           & $2.26\times 10^{5}$
           & $2.18\times 10^{4}$
           & $4.06\times 10^{20}$
         & $\approx$ 700 \\
           C$_8$
           & $7.86\times 10^{5}$
           & $5.24\times 10^{5}$
           & $2.23\times 10^{32}$
         & $\approx$ 2500 \\
\hline
        \end{tabular}
        \caption{
            Wave function size and computational effort for two
        representative calculations C$_n$, where $n$ indicates the number of
        molecules in the cavity. $N_\Psi$ correspond to the total number of
        complex coefficients representing the wave functions, $N_A$ is the size
        of the $A$-vector in Eq.~\ref{eq2:ansatz1} and $N_\textrm{prim}$ is the
        size of the primitive direct product basis, i.e. the size of the
        corresponding standard wave function in a hypothetical numerically exact
        propagation on the full grid. The wall-clock time per propagated fs
    $t_\textrm{CPU}$/$N_\textrm{fs}$ has been scaled to one single processor
        (Intel Xeon E5-2680 v4 @ 2.4 GHz).
Actual calculations were performed using shared-memory parallelization with up
to 28 processors~\cite{bri08:141}.
        }
        \label{tab:mctdh}
    \end{table}

    \subsection{Molecular model}

\begin{figure}[h!]
       \begin{center}
          \includegraphics[width=0.95\columnwidth]{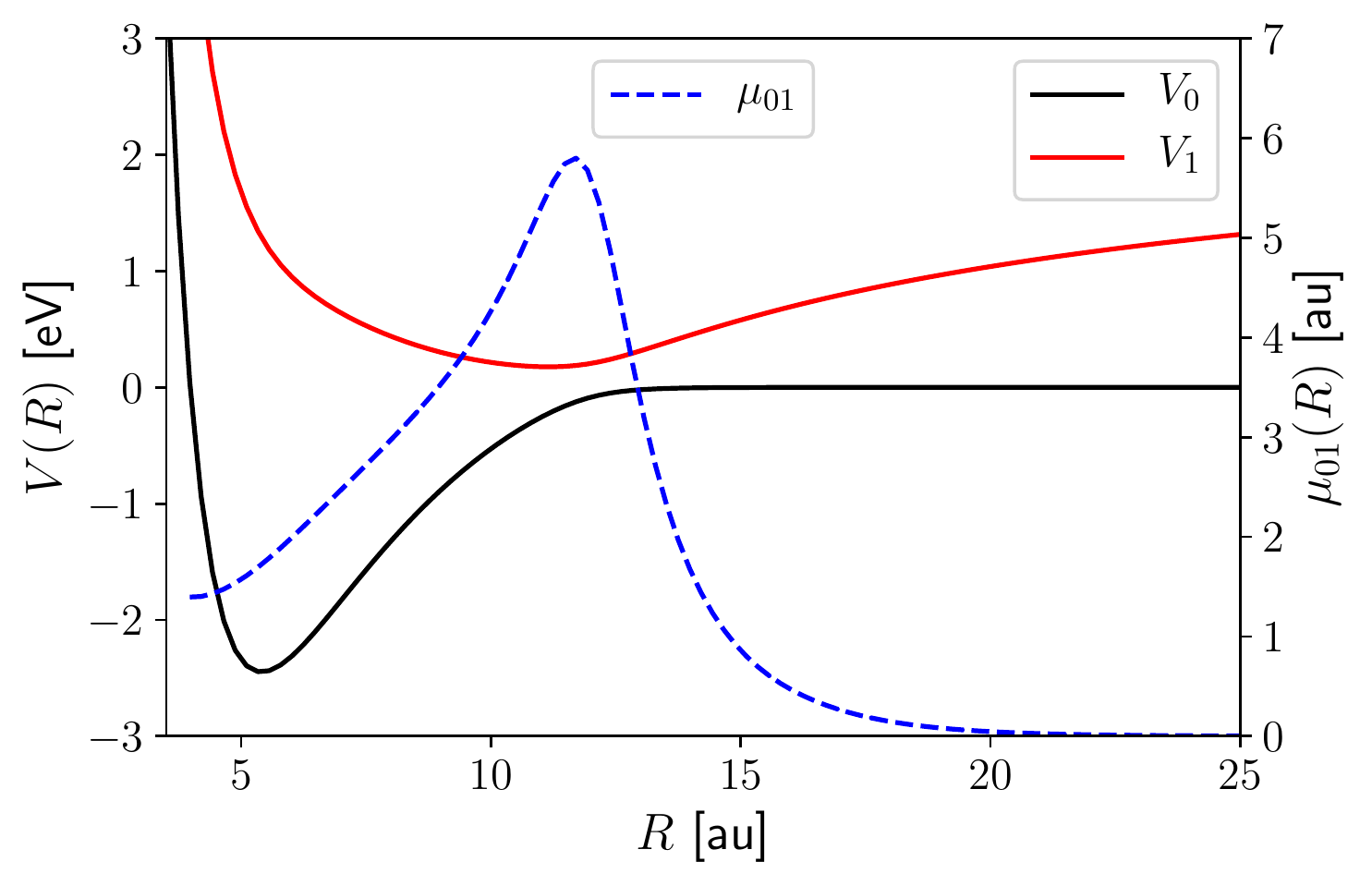}
       \end{center}
        \caption{Potential energy curves of the ground $V_0$ and first singlet
        excited $V_1$ electronic states (solid curves) and transition dipole
        moment $\mu_{01}$ along the molecular
        axis (dashed curve) for the NaI molecule obtained at the CASSCF level of
        theory (See main text for details).}
       \label{fig:potdip}
    \end{figure}

    Following recent theoretical investigations involving one molecule in a
    cavity~\cite{kow16:2050},
    we will consider here an ensemble of NaI molecules. This system is
    well known in the femtochemistry literature and features a relatively simple
    and well understood dynamics upon photo-excitation to its first singlet
    excited electronic state.
    The potential energy surfaces and transition dipole matrix elements are
    shown in Fig.~\ref{fig:potdip}
    and were calculated with the GAMESS-US~\cite{sch93:1347} package
    at the complete active self-consistent field (CASSCF)
    level of theory CAS(8,8), involving 8
    valence orbitals
    (4 occupied, 4 virtual in the
    reference configuration)
    and 8 electrons. The inner valence and core electrons were treated with
    the SBKJC effective core potential~\cite{ste92:612}.

    It is not the goal of this investigation to obtain
    highly accurate potentials for NaI, and the level of electronic theory just
    described is sufficient for the purpose of this work. For example,
    the PES used in Ref.~\cite{kow16:2050} feature a smaller gap between
    the two curves at the avoided crossing.
    While this gap affects non-adiabatic
    transitions within each member of the ensemble, these transitions
    are switched off in
    the present model with the purpose of highlighting the
    dynamical effects induced by the coupling to the cavity.

\section{Results and Discussion}
    \label{sec:results}

    \subsection{General considerations on the cavity-ensemble model}

    Throughout this work
    the effective cavity-matter coupling is taken as $g/\omega_c$=0.01, where
    $g =\lambda\sqrt{\hbar\omega_c/2}$ is the rms vacuum electric field
    amplitude of the cavity mode~\cite{har89:24} and $\lambda$ was introduced
    around Eqs.~(\ref{eq:hcav})-(\ref{eq:hcav2}).
    The use of $g$ as the coupling strength parameter
    is in line with previous investigations~\cite{gal15:41022,gal16:13841,kow16:2050}
    and facilitates direct comparisons.
    This coupling strength
    is small compared to the single-molecule ultra-strong coupling regime,
    characterized by a Rabi splitting of the polaritonic energy levels (at zero
    detuning) $\hbar\omega_R=2 g\mu_{01}$ comparable to the transition energy.
    For the NaI molecule at the equilibrium geometry, the transition dipole
    $\mu_{01}$ (cf. Fig~\ref{fig:potdip}b) is about $1.7$~au, which results in
    $\hbar\omega_R\approx 0.13$~eV.
    For more than one molecule, the collective Rabi splitting becomes
    \mbox{$\hbar\Omega_R=2 g\mu_{01}\sqrt{N}$}~\cite{tho92:1132}.
    Hence for
    $N=5$ (calculations with up to $N=8$ were performed)
    the collective Rabi splitting
    at the Franck-Condon (FC) equilibrium
    geometry $R_{\mathsf{eq}}$ is of the order
    \mbox{$\hbar\Omega_R\approx 0.3$~eV}, in line
    with earlier investigations~\cite{gal15:41022,gal16:13841,kow16:2050}.
    %

    In this work the cavity coupling $g/\omega_c$ is kept \textit{constant} for
    different numbers of molecules $N$ (or molecular density $N/V$) in the
    cavity.  This is in contrast to other works where the collective Rabi
    splitting is kept constant by scaling the cavity coupling by
    $1/\sqrt{N}$~\cite{gal16:13841,gal17:136001}.
    In the present context, Hamiltonian~(\ref{eq:htot})-(\ref{eq:dipmol})
    describes
    the coupling of each individual member of
    the molecular ensemble to a \textit{specific} cavity characterized by
    the quantization volume $V$, which is also the actual volume of the cavity.
    On the other hand, the Rabi splitting is a macroscopic quantity that emerges
    from the microscopic
        interactions.
    Under this consideration there is no \textit{a
    priori} reason to fix $\Omega_R$ for different number of molecules
    $N$.
    %
    Therefore, our focus here will be on collective effects that emerge when
    varying $N$ inside a
    given cavity, always in the limit of negligible direct interaction among
    ensemble members.

        Concluding, we note that the coupling term between laser light and
        the hybrid system in
        Eq.~\ref{eq:hlas} describes the interaction of the laser pulse with the
        molecules only. This is justifiable as long as the laser is non-resonant
        with the cavity frequencies.
        In situations, as discussed below, in which both
        the laser and the cavity are resonant with the Franck-Condon
        transition of the molecules, extra care with this assumption is
        required.
        It can, e.g., be conceived for the
        configuration in
        which the wave vectors $\vec{k}_i$ of the electromagnetic
        modes of the laser lie parallel
        to the plane of the cavity mirrors (in a Fabry-P{\'e}rot configuration),
        such that the external laser field and the cavity modes share
        a common
        polarization axis~\cite{car12:125424}.
        On the other hand, direct coupling between the laser field and the
        cavity modes, for example in open plasmonic structures, may turn out to be
        the dominant coupling mechanism. From a computational perspective, the
        model discussed above may then be easily extended to include the
        corresponding laser-cavity coupling terms obtained either from phenomenological or
        first-principles considerations~\cite{dut05_CQED}.

    \subsection{Femtosecond laser pump of a hybrid ensemble-cavity system}
    \label{sec:laser}

\begin{figure}[h!]
       \begin{center}
          \includegraphics[width=0.95\columnwidth]{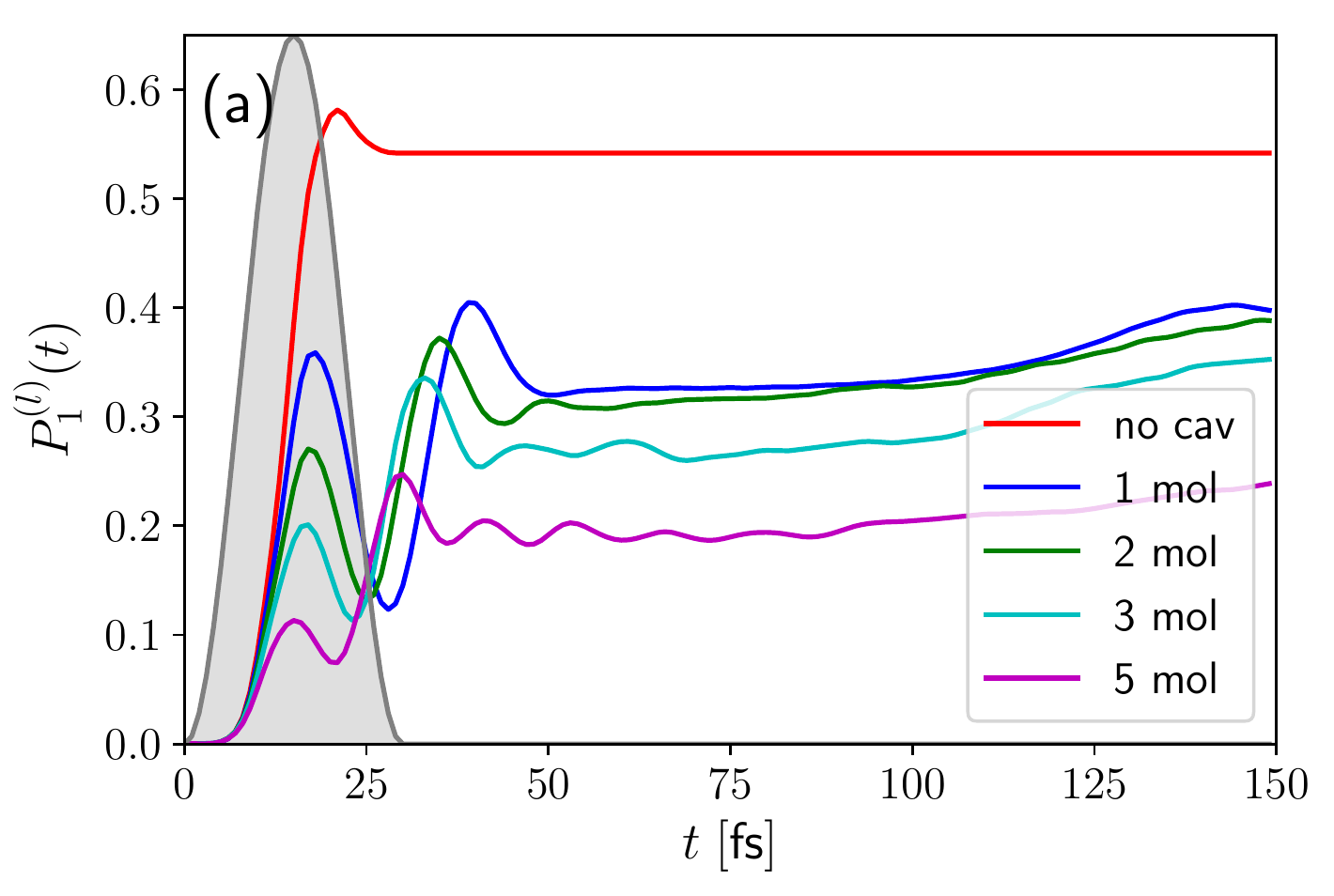}
          \includegraphics[width=0.95\columnwidth]{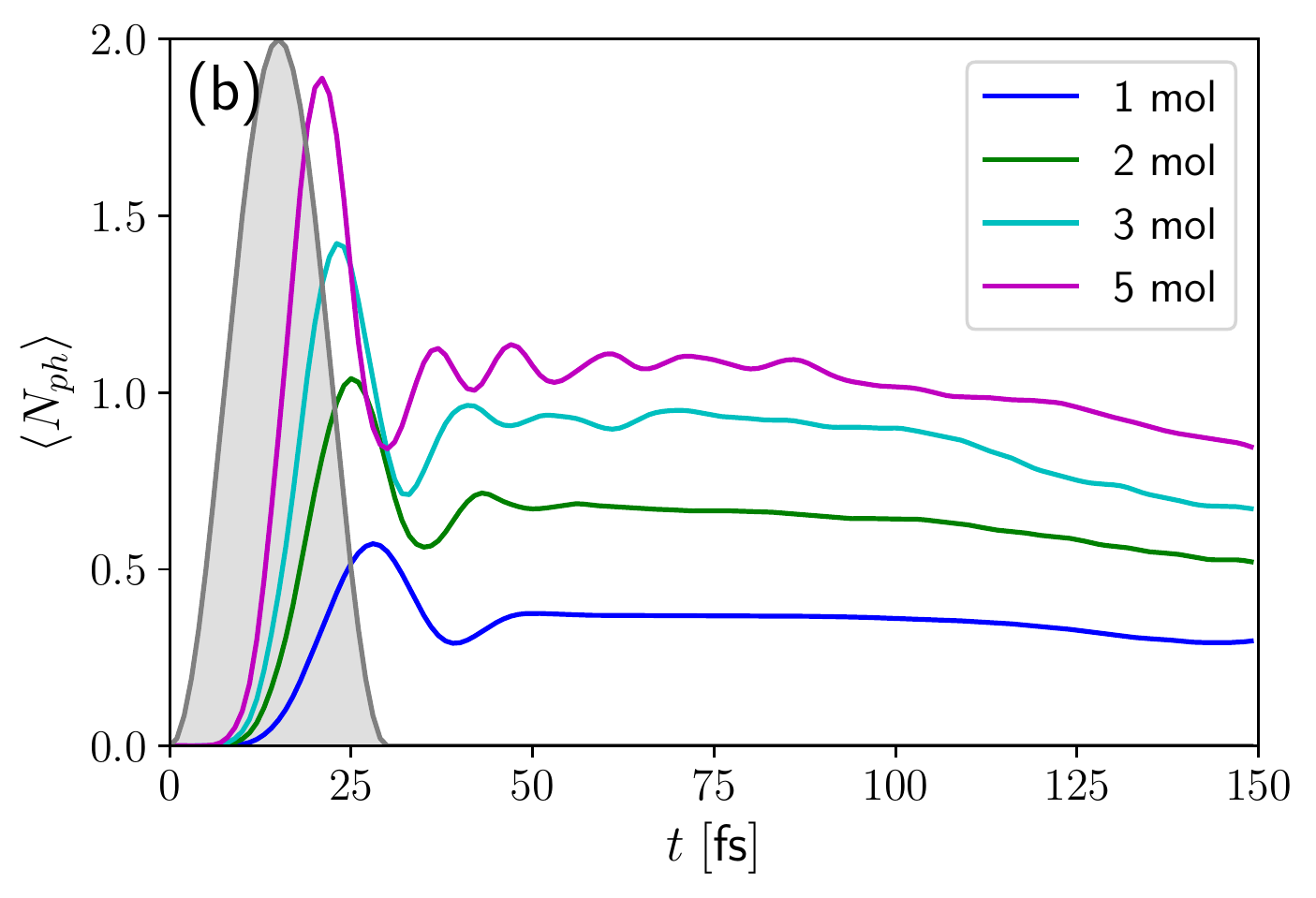}
       \end{center}
       \caption{(a) Single molecule excited state population
                \mbox{$P_1^{(l)}(t)=|\langle \psi_1^{(l)}|\Psi(t)|\rangle|^2$}
                and (b) expectation value of
                the cavity photon number
                \mbox{$\langle N_{ph}\rangle =
                \langle\Psi(t)|\hat{a}^\dagger\hat{a}|\Psi(t)\rangle$}
                as a function of time for propagations without a cavity and for
                1 to 5 molecules in the cavity.
                The grey area indicates the envelope of the laser pulse
                (cf. Eq.~(\ref{eq:efield})).
                The peak laser amplitude is 0.003 au
                (corresponding to $3.16\times 10^{11}$ W/cm$^2$) and
                the photon energy $\hbar\omega_L=3.8$~eV.}
       \label{fig:Pop_Nph_00030}
    \end{figure}

    The excitation of a molecular ensemble with a short
    femtosecond laser pulse is considered first. The laser resonantly
    couples the ground
    and first excited electronic states of each molecule,
    and the cavity is resonant as
    well with this electronic transition at the Franck-Condon geometry.
    %
    The cavity and laser photon energy are both
    $\hbar\omega_c=\hbar\omega_L=3.8$~eV and
    the laser pulse is modelled as
    \begin{align}
        \label{eq:efield}
        E(t) = \sin^2\left( \frac{\pi t}{T_L}\right)\cos(\omega_L t),
    \end{align}
    where $T_L$ is the total pulse duration and the full
    width at half maximum (FWHM) of the amplitude is $\tau_L=T_L/2$.
    Only the first period of the envelope
    function is considered and the pulse has zero amplitude at earlier and later
    times.
    The excitation by the short femtosecond pulse takes place in the impulsive
    regime: the relatively heavy nuclei of NaI practically do not move during
    the pulse duration.
    The most intense pulse considered corresponds to 0.003 au of peak field
    amplitude ($3.16\times 10^{11}$~W/cm$^{2}$). This results in nearly a 50-50
    superposition of the $|\psi_0\rangle$ and $|\psi_1\rangle$ electronic states for
    isolated molecules \textit{without} a cavity, as illustrated in
    Fig.~\ref{fig:Pop_Nph_00030}a. This pulse intensity is far from
    field-ionizing conditions at the optical frequencies considered here.
    %

    The quantities of interest are the single molecule excited state population
    \begin{align}
        \label{eq:population}
        P_1^{(l)}(t)=|\langle \psi_1^{(l)}|\Psi(t)|\rangle|^2
    \end{align}
    and the expectation value of the cavity photon number
    \begin{align}
        \label{eq:photonnum}
        \langle N_{ph}\rangle(t) =
        \langle\Psi(t)|\hat{a}^\dagger\hat{a}|\Psi(t)\rangle.
    \end{align}
    For more than one molecule in the cavity the molecular index $l$
    in Eq.~(\ref{eq:population})
    can take any value as all molecules are identical.
    %
    For all molecule number cases considered, Fig.~\ref{fig:Pop_Nph_00030}a shows
    a fast increase of the molecular excitation
    $P_1^{(l)}(t)$ while the pulse
    intensity is still rising, but the presence of the cavity has a dramatic
    effect on how the pulse energy is redistributed in the system.
    Shortly after the molecular excitation probability peaks,
    the cavity mean photon number $\langle N_{ph}\rangle$
    starts increasing as the molecules transfer energy to
    the cavity
    in the onset of the first Rabi cycle.
    The number of molecules modulates the amplitude of the oscillations, as well
    as their period, as
    seen in
    Fig.~\ref{fig:Pop_Nph_00030}b.
    %
    This ongoing Rabi dynamics
    is of course a consequence of the coherent wave packet prepared by the laser
    pulse.

    For the case of $N=5$, characterized by the largest collective coupling of
    the ensemble and the cavity, an almost direct
    energy transfer from the laser pulse to the cavity mode takes place.
    The population of molecular excitations, which mediate the energy transfer
    to the cavity, remain very low during the duration of the
    laser pulse, which is reminiscent
    of a Stimulated Raman Adiabatic Passage (STIRAP)-type
    mechanism~\cite{tannor-book}.
    These dynamics correspond to a regime dominated by coherent time evolution
    among the polaritonic states impulsively populated by the short
    laser pulse. The time evolution of these oscillations roughly corresponds
    to $\cos(\Omega_{\mathsf{Rabi}} t)$, which would be exactly the case in the simpler
    scenario of two-level atoms coupled to the same cavity and external field.
    Nuclear motion, however, quickly quenches these dynamics by destroying the
    necessary electronic-photonic coherence.
    To see this, one can argue qualitatively for a moment in terms of a simple
    system with electronic $s=\{0,1\}$ and cavity photon $P=\{0,1\}$ degrees of
    freedom and an
    additional nuclear coordinate $R$.
    Before the onset of nuclear displacements the state of the system prepared
    by the laser can be written as
    \begin{align}
        \label{eq:model1}
        |\Psi(t)\rangle
        = \phi(R)
        \Big( c_{01}(t) |\psi_0\rangle |1\rangle
            +  c_{10}(t) |\psi_1\rangle |0\rangle
        \Big)
    \end{align}
    where the off-diagonal coupling between both states is
    $g\mu_{01}(R)$ (cf. also Eq.~(\ref{eq:hamapp})) and
    the evolution of the photon number in the cavity given by
    \begin{align}
        \label{eq:Nph-simple}
        \langle N_{ph}\rangle = |c_{01}(t)|^2 \approx
        \sin^2(g \mu_{01}(R_{\mathsf{eq}})t/\hbar).
    \end{align}
    %
    After the nuclear wave packet in the excited electronic state
    leaves the FC region the state of the
    system can qualitatively be written as
    \begin{align}
        \label{eq:model2}
        |\Psi(t)\rangle
         =
        c_{01}(t) \phi_0(R,t) |\psi_0\rangle|1\rangle
         +
        c_{10}(t) \phi_1(R,t) |\psi_1\rangle|0\rangle,
    \end{align}
    where the nuclear wave packet evolves differently in the two molecular
    electronic states and
    where both $\phi_s(R,t)$ are normalized.
    Now, as the nuclear wave packet $\phi_1(R,t)$ in the excited electronic state
    becomes non-resonant with the cavity mode, the coupling to the cavity
    vanishes and $|c_{01}(t)|$ becomes almost constant.
    The large oscillations in
    photon number stop.
    However, the $\phi_0(R,t) |\psi_0\rangle|1\rangle$ component
    corresponds to the nuclei
    still at the FC region ($R\approx R_{\mathsf{eq}}$)
    with one cavity photon. Therefore, the cavity
    keeps promoting the molecular system from its ground to its excited electronic state,
    where it
    dissociates and cannot emit a photon back to the cavity. These dynamics
    continue until the cavity is completely relaxed. The onset of this
    dynamical regime is indeed
    seen in the MCTDH numerical results in Fig.~\ref{fig:Pop_Nph_00030}b
    for all molecular numbers as a continuous decrease in photon number that
    starts at 40 to 50~fs and has its counterpart in an increase of the
    excitation probability per molecule in Fig.~\ref{fig:Pop_Nph_00030}a.

    \begin{figure}[h!]
       \begin{center}
          \includegraphics[width=0.95\columnwidth]{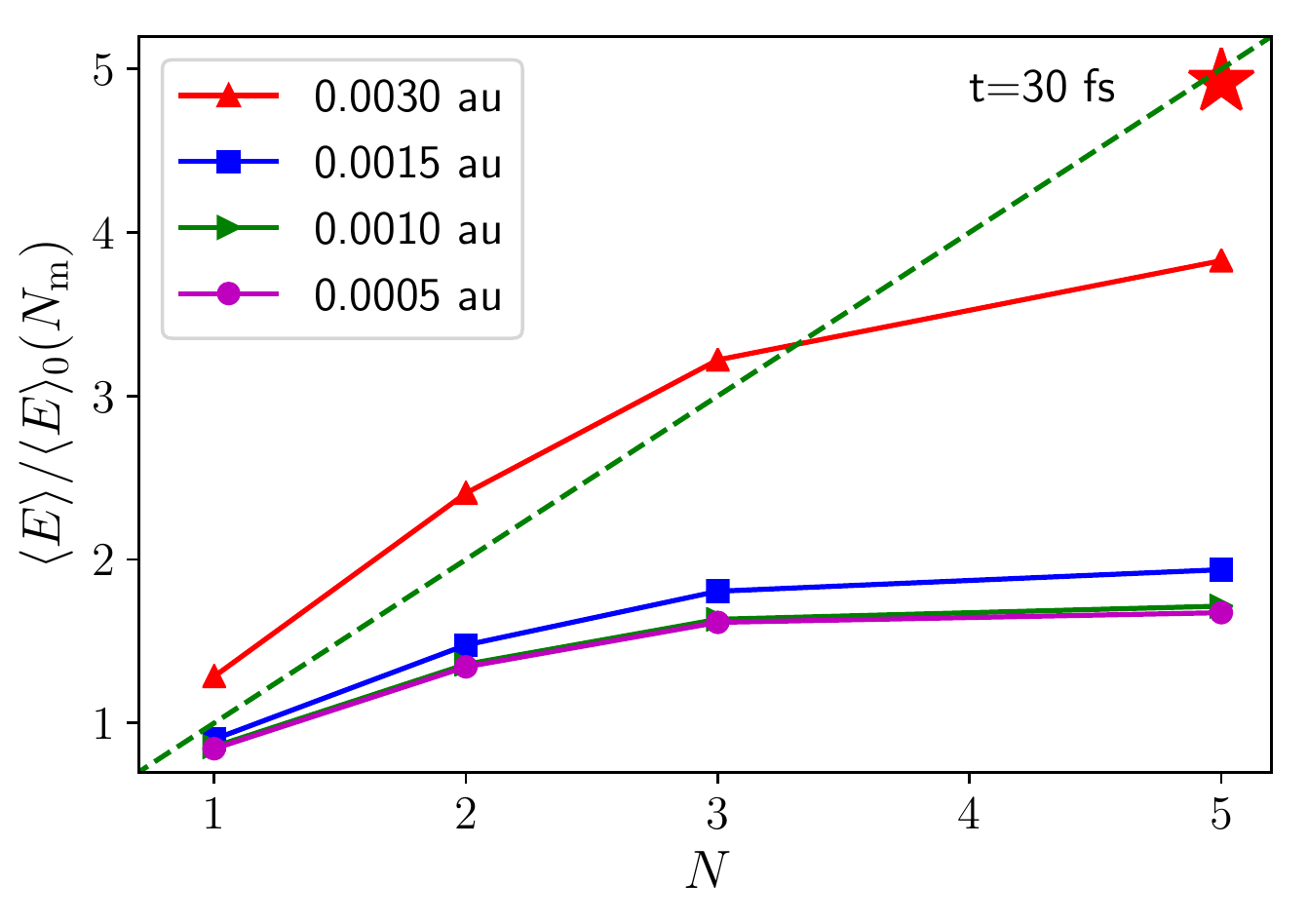}
       \end{center}
       \caption{Energy transfer to the hybrid ensemble-cavity system relative to
        the energy transfer to a single isolated molecule for different molecule
        numbers and laser peak intensities after the laser pulse is over.
        The laser pulse has the intensity
        profile shown in Fig.~\ref{fig:Pop_Nph_00030} except for the simulation
        with 5 molecules and marked with a read star, in which the pulse energy
        is preserved but the FWHM of the intensity profile is shortened by a
        factor of 3 to increase the laser bandwidth.}
       \label{fig:EF}
    \end{figure}
    We just discussed
    how the onset of nuclear motion strongly modifies and quenches
    the coherent polaritonic dynamics that a femtosecond laser imprints onto
    the hybrid system.
    We see as well that $P_1^{(l)}(t)$
    in Fig.~\ref{fig:Pop_Nph_00030}a
    reaches a smaller value after the pulse
    for $N=5$ as compared to
    $N=1$.
    Although $\langle N_{ph}\rangle$
    in Fig.~\ref{fig:Pop_Nph_00030}b
    grows faster and
    reaches a larger value after the pulse for $N=5$ molecules, this does not
    compensate for the much lower molecular excitation.
    Indeed, the total amount of energy
    \textit{per molecule} transferred by the external laser to the system is
    decreasing with the number of molecules, as illustrated in
    Fig.~\ref{fig:EF} for different laser intensities. This figure presents the
    ratio between the total energy transferred to the hybrid system by the laser
    pulse and the total energy transferred to a single molecule with no cavity,
    also after the pulse.
    The results in Fig.~\ref{fig:EF} illustrate how,
    for increasing number of molecules and equal laser
    conditions, the relative likelihood that one of the molecules is
    photoexcited and is able to start a
    photochemical process decreases, which is reminiscent of the findings in
    Ref.~\cite{gal16:13841}.

    To understand the origin of this trend we note that
    the collective Rabi splitting at the FC point
    reaches a value $\Omega_R\approx 0.3$~eV for $N=5$, whereas a pulse
    duration 15~fs FWHM corresponds to a bandwidth of about 0.27~eV.
    Therefore, as the number of molecules increases, the
    upper and lower polaritonic states of the hybrid system
    drift out of resonance
    with the laser pulse.
    To demonstrate that energy transfer to the coupled ensemble is
    determined by the bandwidth of the femtosecond pulse,
    a simulation with a shorter pulse of duration
    $\tau_S=\tau_L/3$ and $N=5$ is made. The maximum field amplitude
    is scaled by a factor
    $\sqrt{3}$ to maintain the area under the intensity
    envelope of the two pulses exactly constant, therefore achieving the same pulse
    energy. The pulse bandwidth increases now to about 0.8~eV.
    The final energy ratio for this simulation is shown with a star in
    Fig.~\ref{fig:EF} and falls almost exactly on the
    linear trend as a function of the number of molecules.
    It is emphasized that the denominator $\langle E\rangle_0$ is the same
    for the long and short pulse simulations, namely the energy transferred
    by the longer pulse to a single isolated molecule.

    Therefore, if the pump laser pulse is sufficiently short to cover with
    its bandwith the splitted polaritonic states in the cavity, the energy
    transferred per molecule becomes the same as for a single isolated molecule
    with that same laser.
    From a purely time-dependent perspective, once the pump laser is
    significantly shorter than the Rabi period, the molecules in the cavity
    become excited by the laser before being able to interact with the cavity and during
    these initial moments it is as if the cavity would not be present.

    \begin{figure}[h!]
       \begin{center}
          \includegraphics[width=0.95\columnwidth]{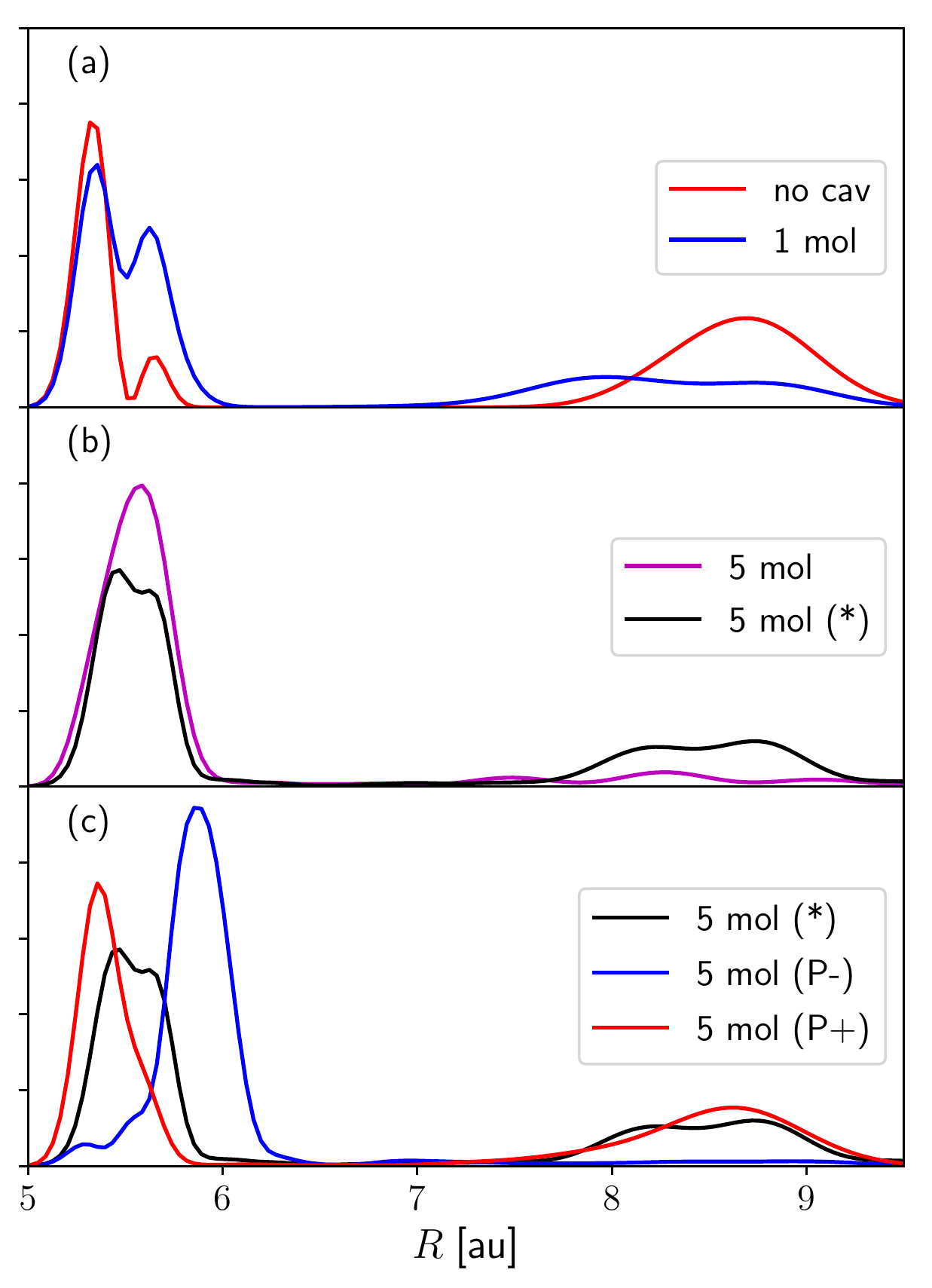}
       \end{center}
       \caption{Probability
        density $\rho(R_l)$ for the nuclear degrees of freedom at
        $t=100$~fs for the same laser conditions as in
        Fig.~\ref{fig:Pop_Nph_00030}.
        (a) no cavity and one molecule in the cavity.
        (b) 5 molecules in the cavity. $(*)$ indicates an increased laser
        bandwidth.
        (c) 5 molecules in the cavity. $(*)$ as in (b).
        $(P-)$ and $P(+)$ indicate laser tuned to the lower and higher polariton
        branches, respectively.
        (see text for details).}
       \label{fig:rhoR}
    \end{figure}
    A consequence of this is that the probability for the
    photochemical process in the excited electronic state to take
    place becomes independent
    of the number of molecules in the cavity.
    This last point is further illustrated in Fig.~\ref{fig:rhoR}, where the
    probability density $\rho(R_l)$ for the nuclear degrees of freedom is shown
    at $t=100$~fs.
    For one single molecule in the cavity, the area under the curve at the right side of the coordinate
    for the dissociating systems is essentially the same as for one molecule
    without a cavity, whereas it is strongly suppressed
    for 5 molecules,
    as seen in Fig.~\ref{fig:rhoR}b,
    due to the lower level of excitation per molecule.
    Figure~\ref{fig:rhoR}b also illustrates, however, how the propagation initiated with a
    larger laser bandwidth restores the dissociation probability.
    %
    %

    This last point
    contrasts with Ref.~\cite{gal16:13841}, where the
    photochemical reaction likelihood was found to decrease with increased
    number of
    molecules in a cavity.  There, the trigger process was not
    considered and the molecules were directly promoted to the lower adiabatic
    polaritonic PES.
    To shed light on those aspects, two further calculations using the
    longer laser pulse with photon energy shifted by $\pm 0.2$~eV are
    considered. These target, respectively, the lower and upper polariton
    branches.
    As the laser is tuned to the lower branch, indicated as $(P-)$ in
    Fig.~\ref{fig:rhoR}c, the reaction is almost completely suppressed.
    However,
    the energy transfer to the hybrid system
    relative to a single molecule is now
    $\langle E \rangle/\langle E_0 \rangle=7.28$ (cf. Fig.~\ref{fig:EF}),
    even larger than the value of
    $\approx 5$ corresponding to a broadband, shorter excitation and
    signaled with a red star in Fig.~\ref{fig:EF}.
    So, even if a higher excitation density is achieved by tuning to the lower
    polariton, presumably the large admixture of ground electronic state character
    in the lower branch-states
    suppresses the start of the photochemical process,
    in line with the findings in Ref.~\cite{gal16:13841}.

    On the other hand, as seen in Fig.~\ref{fig:rhoR}c,
    when the laser is tuned to the upper polariton branch,
    the opposite effect is achieved and the photodissociation
    of each ensemble member is enhanced to the level
    of the isolated molecule
    case, which is explained by the enhanced excited
    electronic state character of the upper polaritonic branch.
    Therefore, a description of the pump process and of the
    initially prepared state are in general crucial elements to
    predict the reaction mechanisms in the coupled ensemble-cavity
    system.

    \subsection{Superradiant energy transfer to the cavity}
    \label{sec:coherence}

    As discussed before,
    a short femtosecond pulse imprints a coherent excitation
    on the coupled ensemble-cavity system.
    We now focus on the coherent dynamical evolution of the coupled system and the
    participation of the cavity mode in the dynamics of the individual
    molecules.
    \begin{figure}[t!]
       \begin{center}
          \includegraphics[width=0.95\columnwidth]{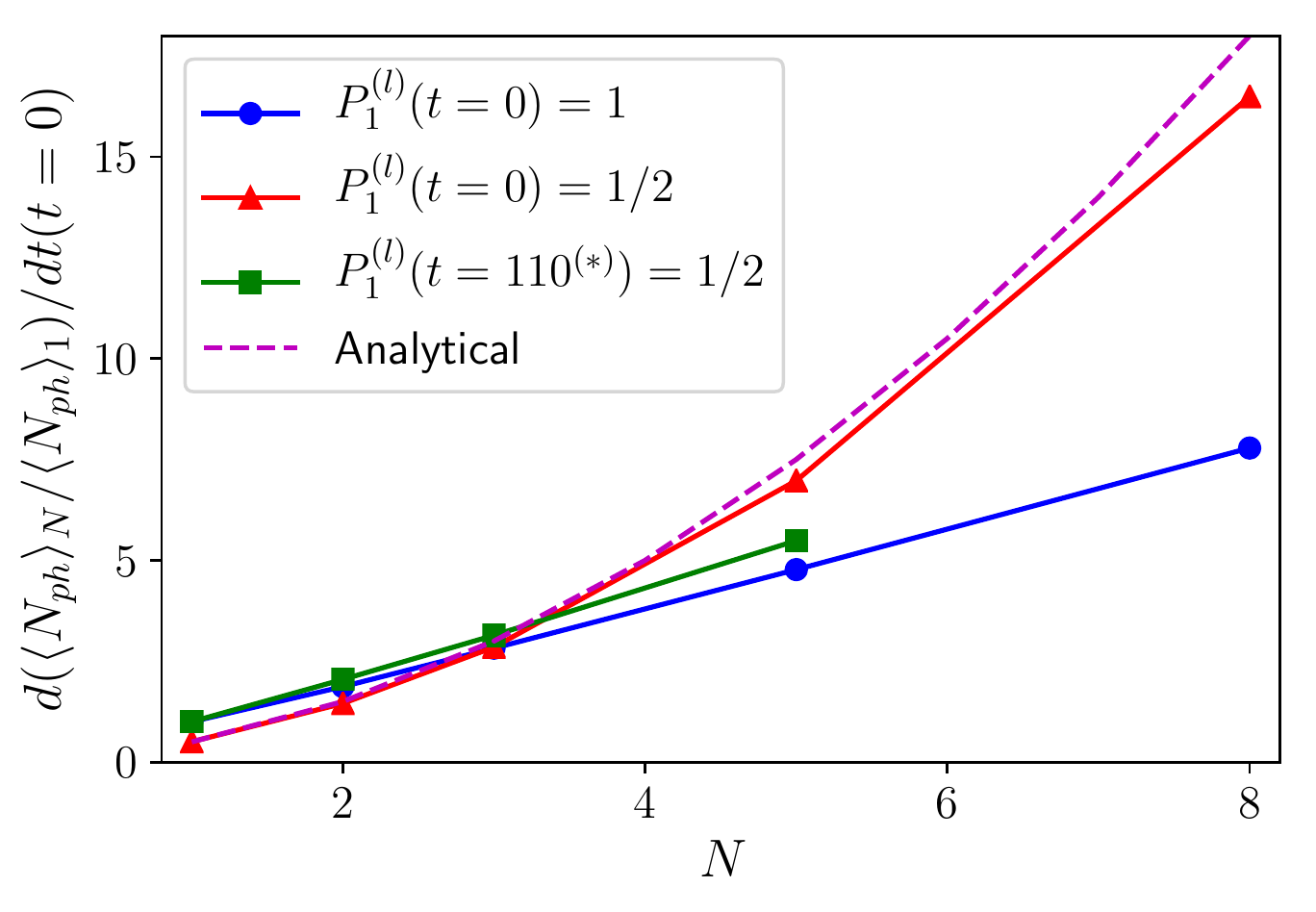}
       \end{center}
       \caption{Relative energy transfer rate to the cavity mode relative to the
        single molecule case. All molecules excited with
        probibility one (blue circles) and all molecules excited with probability
        $1/2$ to the excited state (red triangles),
        immediately after excitation and for a cavity resonant with the FC
        transition energy.
        Relative rate of energy transfer
        for excited molecules in the $V_1$ PES at $t=110$~fs, as they become
        resonant with a cavity $\omega_c=1$~eV (green squares).
        The analytical curve corresponds to the term in parenthesis in
        Eq.~(\ref{eq:pnum2nd}) for $p_0=p_1=1/2$ and $S_{01}=1$.}
       \label{fig:rad_rate}
    \end{figure}
    The cavity is still tuned to be resonant at the FC geometry,
    and it is assumed
    that the molecules have been coherently excited by a laser pulse
    substantially shorter than their collective Rabi period
    in the cavity. The initial state at $t=0$ thus reads
    \begin{align}
        \label{eq:istate}
        |\Psi(0)\rangle =
        \prod_{l=1}^{N}
        \left(
            \sqrt{p_0} \phi_0^{(l)}(R_l) |\psi_0^{(l)}\rangle
          + \sqrt{p_1} \phi_1^{(l)}(R_l) |\psi_1^{(l)}\rangle
        \right) |0\rangle
    \end{align}
    where the same definitions as in Eq.~(\ref{eq:wfn0}) are used.
    $\phi_s^{(l)}(R_l)$ is the nuclear wavepacket evolving on the PES
    of electronic state
    $|\psi_s^{(l)}\rangle$, and $p_{s}$ is the initial electronic population
    of this state.
    In contrast to Eq.~(\ref{eq:wfn0}), the norm of the nuclear
    wave packets, which is the population of the corresponding electronic state,
    has been singled out for the sake of clarity, and hence the $\phi_s^{(l)}(R_l)$
    are assumed normalized.
    Equation~(\ref{eq:istate}) is a single Hartree product of the form of
    Eq.~(\ref{eq:wfn0}) and corresponds to the ground state of the complete
    system for $p_1=0$.

    The total energy transferred from the molecules to the
    cavity at time $t$ is proportional to the number of cavity
    photons $\langle N_{ph}\rangle(t)$.
    In second order in time the number of cavity photons is
    given by
    \begin{align}
        \label{eq:pnum2nd}
        \langle N_{ph} \rangle(t)
        & = t^2 \mu_{01}^2 g^2 \bigg(
        N \,p_1 + \left(N^2 -N\right) p_0 p_1 |S_{01}|^2 \bigg),
    \end{align}
    where the nuclear overlap $S_{01}=\langle \phi_0^{(l)} | \phi_1^{(l)}\rangle$
    is $1$ at short times for a FC transition. The molecular index drops from
    $S_{01}$ because all molecules are assumed to be identical.
    The justification of Eq.~(\ref{eq:pnum2nd}) is given in the Appendix.
    It is interesting to note that, if the nuclear overlap is set to one as for
    atomic radiators, the term in parenthesis in Eq.~(\ref{eq:pnum2nd}) is the
    same as the linear (in time)
    radiation rate for an ensemble of coherently excited
    radiators coupled to an electromagnetic mode, which was
    first described by Dicke and termed superradiance~\cite{dic54:99}.
    The quadratic scaling with time in Eq.~(\ref{eq:pnum2nd}) is a natural
    consequence of Schrödinger's equation at short times and is of no particular
    interest.
    We
    are interested, instead, in the scaling with the molecular number $N$,
    which varies
    from linear to quadratic and which, as mentioned, is found in
    identical form in Dicke's radiation rate derived using Fermi's golden rule.

\begin{figure}[t!]
       \begin{center}
          \includegraphics[width=0.95\columnwidth]{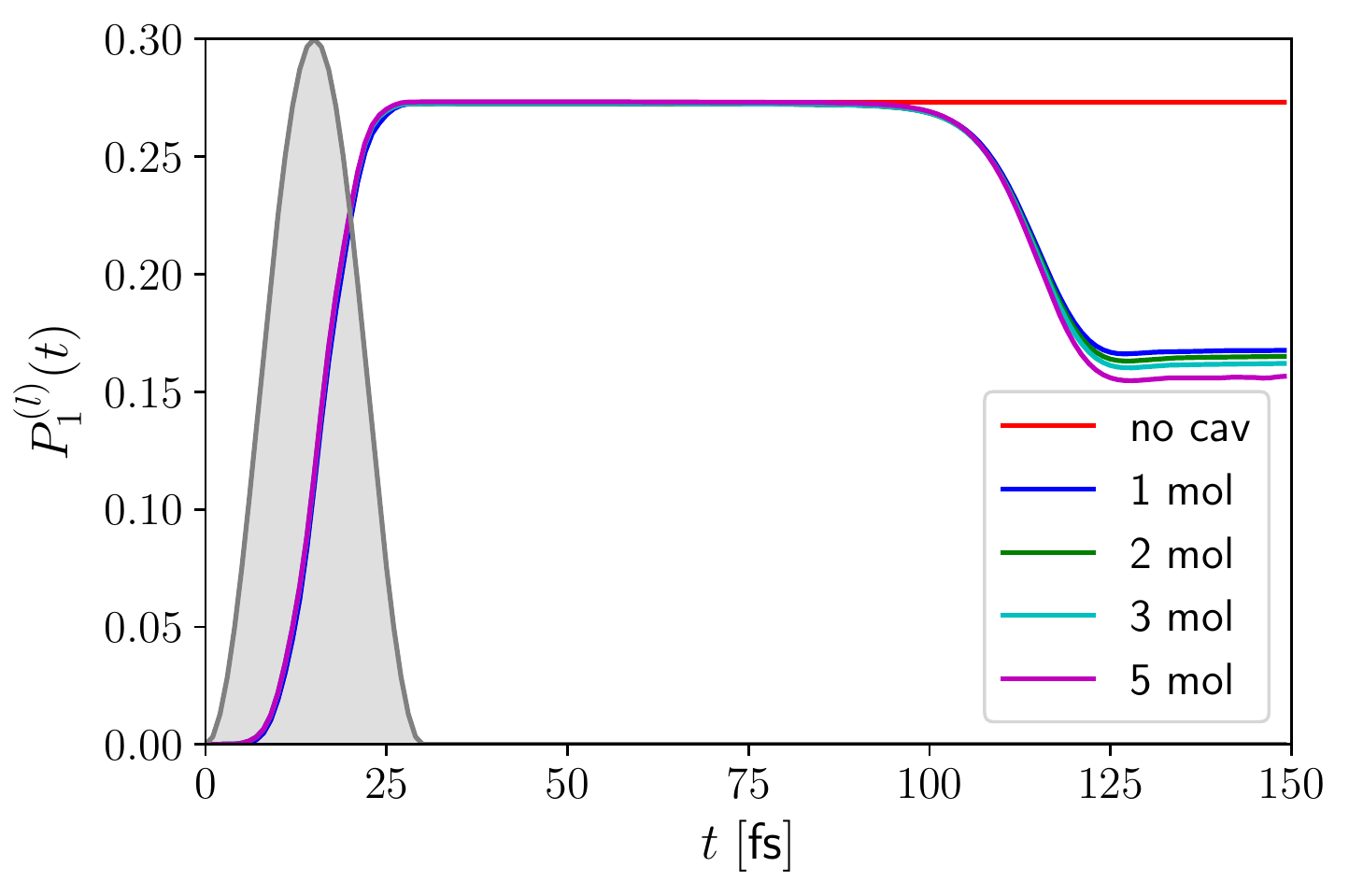}
       \end{center}
        \caption{Single molecule excited state population $P_1(t)$
        for one isolated
        molecule and up to
        5 molecules in the cavity. The pump laser intensity profile is shown
        in grey. The photon energy of the cavity mode is $\hbar\omega_c=1$~eV
        and is resonant with the molecules at an internuclear distance
        $R\approx 10$~au.}
       \label{fig:C1eV_Pop}
    \end{figure}

    When all molecules are promoted to their corresponding state $|\psi_1\rangle$
    with probability $p_1=1$ the energy transfer to the cavity grows
    linearly with the number of molecules, as predicted by
    Eq.~(\ref{eq:pnum2nd}) and numerically shown in
    Fig.~\ref{fig:rad_rate}.
    In contrast, when molecules
    are promoted to
    a coherent superposition of the ground and
    excited electronic states with \mbox{$p_0=p_1=1/2$} the energy
    transfer to the cavity scales quadratically with
    $N$.
    %
    However, as
    the nuclear overlap drops to zero,
    the coherent quadratic contribution
    disappears and the scaling becomes linear in $N$ again.
    The latter regime is illustrated in Fig.~\ref{fig:C1eV_Pop} for the case
    in which the cavity is resonant with the molecules
    outside the vicinity of the FC point.
    In this set of calculations the cavity
    is resonant with a potential energy gap of
    1~eV, which is reached by the
    molecules roughly 110~fs after photoexcitation.
    The population transfer \textit{per molecule} back to the ground
    electronic state at $t\approx 110$~fs
    is practically independent of $N$, and the number of cavity photons
    $\langle N_{ph}\rangle$ (not shown) after the molecules have passed by
    the interaction region is just proportional to $N$.
    This is illustrated with the square marks in
    Fig.~\ref{fig:rad_rate}.
    The return to a linear scaling for the energy transfer to the cavity
    with $N$ outside
    the FC region is hence the consequence of the loss of wave packet overlap
    between wave packets evolving on different
    PES of the same molecule,
    which effectively results in electronic decoherence.

    Indeed, a simplified expression for the off diagonal element of the reduced electronic density
    matrix of the $l$-th molecule can be derived from Eq.~(\ref{eq:istate})
    assuming that the time-evolved wave packet retains a product structure
    \begin{align}
        \label{eq:rho01}
        \rho_{01}^{(l)}(t) &= \langle\Psi(t)
                                |\psi_0\rangle
                                \langle \psi_1|
                                \Psi(t)\rangle \\\nonumber
                             &= \langle \phi_0^{(l)}(R_l,t)
                                | \phi_1^{(l)}(R_l,t) \rangle
                                \sqrt{p_0}\sqrt{p_1} \\\nonumber
                             &= S_{01} \sqrt{p_0}\sqrt{p_1}.
    \end{align}
    This experession is is valid at short times
    and
    the molecular index has again disappeared in the last line
    because all molecules
    are considered identical.
    Comparison of Eqs.~(\ref{eq:pnum2nd}) and (\ref{eq:rho01}) shows that
    the loss of electronic quantum coherence at the single molecule level
    destroys superradiant energy transfer to the cavity.
    In the case of NaI, with a dissociative excited PES,
    $\rho_{01}^{(l)}(t)=0$ for $t>0$ because of the loss of nuclear wavepacket overlap in
    the second line of Eq.~(\ref{eq:rho01}).
    A more involved expression for $\rho_{01}^{(l)}(t)$ can be obtained from
    the general ansatzt~(\ref{eq:wfn0}), which however does not add any new physical insight to
    the conclusions reached from Eq.~(\ref{eq:rho01}).


    \subsection{Stimulated emission by the cavity}
    \label{sec:photons}

    %

\begin{figure}[h!]
       \begin{center}
          \includegraphics[width=0.95\columnwidth]{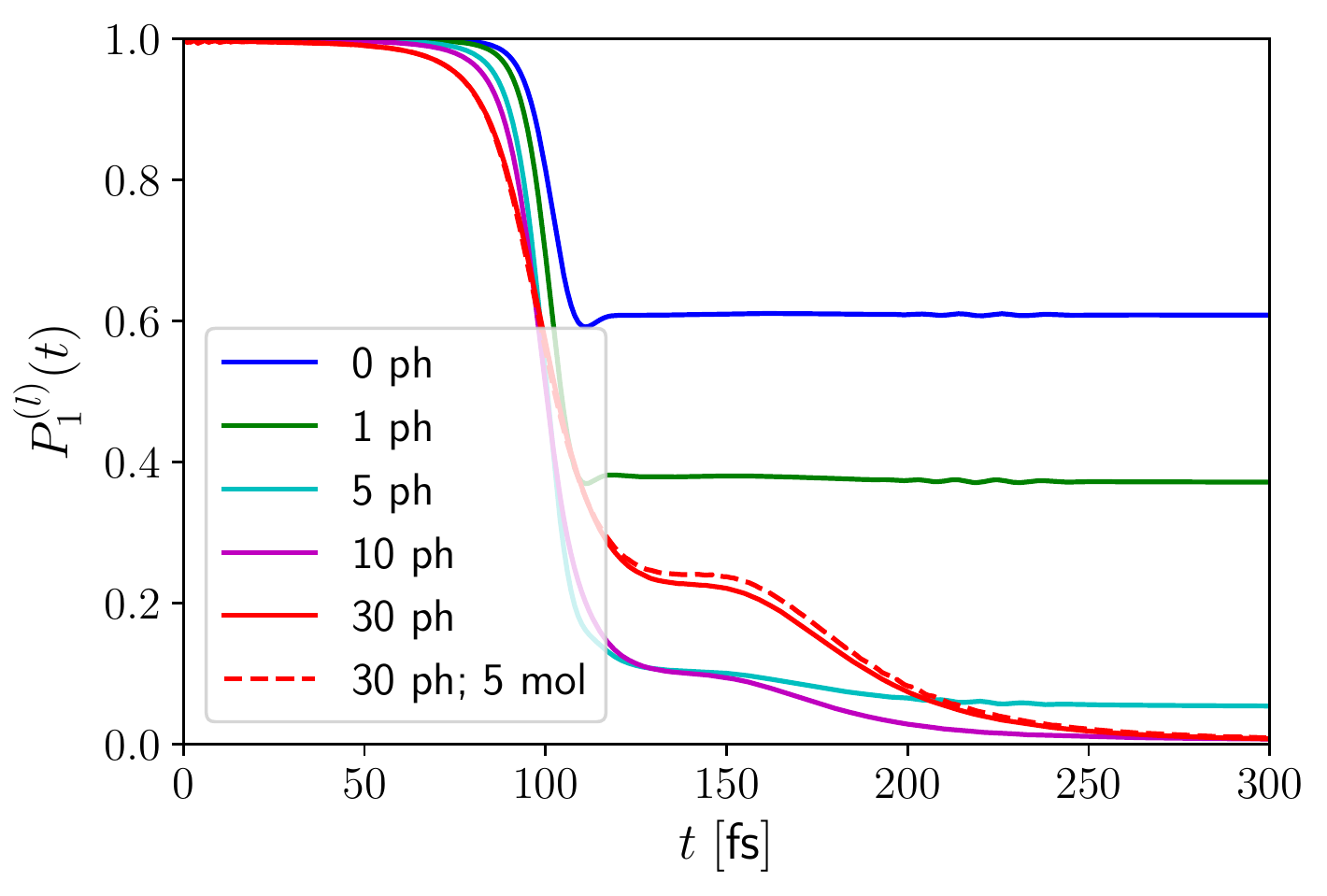}
       \end{center}
       \caption{Single molecule excited state population $P_1(t)$ as a function
        of time. The cavity mode has photon energy $\hbar\omega_c=1$~eV and
        various initial photon numbers are considered. All curves correspond to
        one single molecule in the cavity except for the curve (dashed red)
        corresponding to 30 photons and 5 molecules.
        }
       \label{fig:C1eV_Nph}
    \end{figure}

    The scenario in which the cavity becomes resonant with the molecular
    system
    along a reaction coordinate may be used to steer or
    probe photochemical reactions in ways analogous to the action of a laser
    pulse delayed with respect to the reaction trigger.
    When a photoexcited molecule reaches resonant configurations with the
    cavity, the cavity can stimulate the emission of one photon and dump the
    molecular system to the ground electronic state, as
    shown in Fig.~\ref{fig:C1eV_Pop}, and which is part of the
    molecular relaxation mechanism discussed in~\cite{gal17:136001}.
    An
    obvious control knob in a cavity is presented by its excitation level,
    the number of photons in the
    cavity, since the coupling term proportional to
    $(\hat{a}^\dagger+\hat{a})$ in Eq.~(\ref{eq:hcav2})
    effectively scales with $\sqrt{N_{ph}}$ when applied to cavity state
    $|N_{ph}\rangle$.
    The excited electronic state population for one and five molecules interacting
    with a cavity is shown in Fig.~\ref{fig:C1eV_Nph} for different numbers of
    photons ranging from 0 to 30. The molecules are initially
    prepared in their excited electronic state with unit probability and the
    cavity's photon energy is $\hbar\omega_c=1$~eV.
    The stimulated emission down to the ground electronic
    state strongly depends on the initial photon number of the cavity.
    After going through a maximum
    population dump for 5 to 10 photons, stimulated emission becomes
    indeed less efficient for 30
    photons. The time evolution becomes non-trivial with
    a first plateau and a final
    complete dump, which is indicative of dynamics that proceed on
    effective potential surfaces strongly modified by the cavity coupling.
    We compare now the cases with one and five
    molecules evolving in the
    cavity pre-loaded with 30 photons. It is found that the electronic population
    dynamics of each single molecule in the 5-molecule ensemble
    is essentially the same as the population
    dynamics of the single molecule case, as seen by comparing the two red curves
    in Fig.~\ref{fig:C1eV_Nph}.
    An important difference is that the final state of the
    cavity (not shown) in the latter case corresponds to
    $\langle N_{ph}\rangle=35$ and in the former case to $\langle
    N_{ph}\rangle=31$.

\begin{figure}[t!]
       \begin{center}
          \includegraphics[width=0.95\columnwidth]{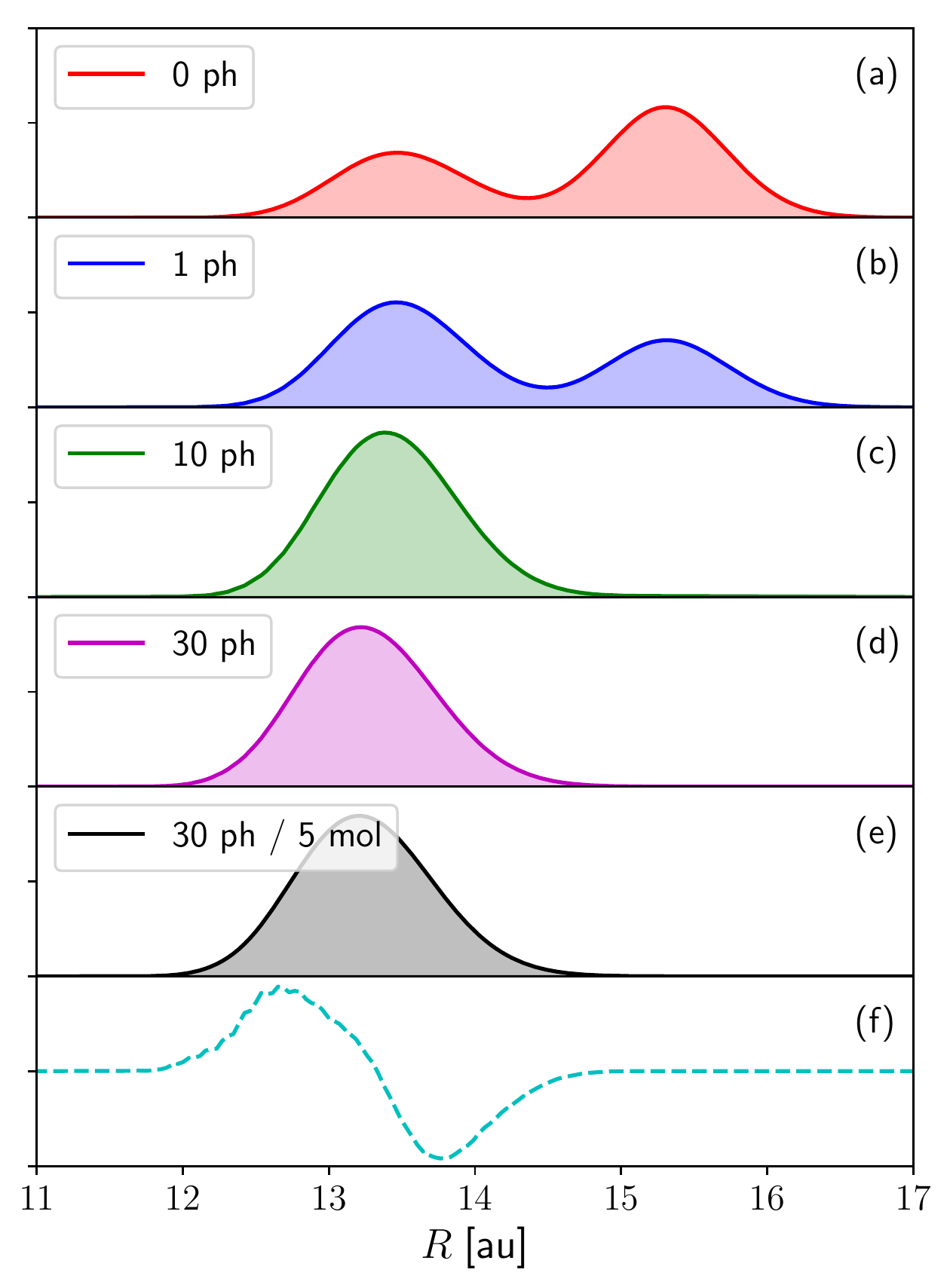}
       \end{center}
        \caption{(a-e) One dimensional probability density $\rho(R,t)$ for
        $t=200$~fs for the same simulations as shown in
        Fig.~\ref{fig:C1eV_Nph}. (f) Probability density difference
        between cases (e) and (d) magnified $\times$ 25.}
       \label{fig:C1eV_WP}
    \end{figure}

    Comparing the probability density $\rho(R,t)$ in Fig.~\ref{fig:C1eV_WP}
    for either one or five molecules
    in the cavity, and after the passage through
    the resonant region at $t=200$~fs, two different regimes are seen:
    first, the
    wave packet splits between excited and ground state components
    for 0 and 1 photons in the cavity.
    Second, the wave packet is almost completely dumped to the ground state
    and remains
    uni-modal for 5 photons and beyond.
    Also for the nuclear evolution, the difference between one and five molecules
    in the cavity becomes insignificant (cf. Fig.~\ref{fig:C1eV_WP}f).
    For a high number of photons, which increases
    the effective molecule-cavity
    coupling, the nuclear wave packet becomes slowed down
    and partially trapped in the region in which the molecular electronic
    energy gap becomes resonant with the cavity. This effect has been seen in
    simulations with one photon but higher coupling
    strengths~\cite{gal15:41022,kow16:2050}.

    The comparison between one and five molecules with 30 cavity photons
    is in line with the discussion in Sec.~\ref{sec:coherence}.
    As the excited molecules reach the
    interaction region there is no nuclear overlap and hence,
    as indicated in Eq.~(\ref{eq:pnum2nd}), the energy
    transfer to the cavity occurs linearly with the number of molecules, i.e.
    each molecule behaves as if it were alone in the cavity.
    Hence, the extent and rate of stimulated emission induced by a cavity
    resonant with the molecules along some photochemical reaction pathway is much
    more dependent on $\langle N_{ph}\rangle$, which determines the
    individual coupling of each molecule to the cavity, than the molecular
    number (or density), which determines the total Rabi splitting but has a
    much smaller effect on the individual dynamics of the ensemble members.
    Moreover, even if a coherent superposition of molecular electronic states
    would be present,
    the number of photons in the cavity cannot be used to alter coherent effects
    that scale with $N^2$,
    as discussed in the Appendix around Eq.~\ref{eq:pnum3}.




    \section{Summary and Conclusions}
    \label{sec:conclusions}

    The real-time dynamics of an ensemble of molecules coupled to an
    electromagnetic cavity mode and pumped by a short femtosecond laser pulse
    was described by quantum wave packet simulations.
    In doing so, all coordinates of the microscopic Hamiltonian of
    the hybrid system were dynamically considered
    without resorting to an adiabatic separation in terms of polaritonic
    potentials and their couplings.

    For a laser pulse with a duration FWHM of 15~fs,
    of order of the collective Rabi period of the hybrid system,
    the short time impulsive dynamics
    is triggered by the laser. These dynamics consist of Rabi oscillations in which
    the overall excitation oscillates between the molecules and the cavity.
    Nuclear motion of the excited molecules leads to electronic-photonic
    decoherence among
    polaritonic states due to loss of nuclear wave packet overlap and
    quickly quenches these dynamics.

    For the same laser conditions, the increase in the number of molecules (or
    the molecular density) leads to a decrease of the energy \textit{per
    molecule} pumped
    to the cavity by the laser pulse, with the corresponding reduction of the
    probability per molecule that the photochemical reaction is started.
    It is found that this effect is caused by the laser drifting out of
    resonance with the bright polaritonic states as the collective Rabi
    splitting increases with the number of molecules.
    Increase of the laser bandwith above the Rabi splitting \textit{for the same
    pulse energy} restores the total energy transferred to linear scaling with
    the number of molecules, and restores the probability that the photochemical
    reaction is triggered by the femtosecond laser pulse.
    Alternatively, tuning the laser to the lower polariton branch leads to a
    suppression of the photochemical process due to the admixture of ground
    state character of such states, in line with Ref.~\cite{gal16:13841}.

    When all molecules are coherently excited to a superposition of their ground
    and excited states, the energy transfer to the cavity occurs in a
    superradiant regime with quadratic scaling with the number of molecules.
    Again, this regime is destroyed by the loss of nuclear wave packet overlap
    within each individual molecule for
    the involved electronic states.
    This also means that the
    probability of photon emission to the cavity mode scales linearly
    with the number of molecules when the photoexcited
    molecules arrive at geometries resonant with the cavity
    along some reaction coordinate.
    The number of photons in the cavity
    has a very noticeable
    effect in stimulating emission to the cavity,
    but cannot be used to alter the
    dynamics related to coherent superposition of molecular electronic
    states.

    In this application we chose NaI as a well known
    molecular system that has been
    the subject of numerous investigations, both experimental and theoretical,
    in the field of femtochemistry, and which undergoes a
    simple photodissociation process upon light absorption. This allowed us
    to concentrate on aspects of the excitation process and energy transfer
    dynamics in the cavity.
    By combining the same basic principles of this work with the multilayer MCTDH method, the
    study of a larger number of (multi-dimensional) molecules in a
    cavity, including possibly local dissipative baths and, if required,
    a larger number of cavity modes,
    is well within reach and opens new avenues for the rigorous investigation of
    cavity femtochemistry problems.


\section{Acknowledgments}

    I am grateful to Prof. Lars B. Madsen for stimulating discussions and
    critical reading of this manuscript.
    I wish to dedicate this work to Prof. Hans-Dieter Meyer on occasion of his
    70th birthday, and to thank him for his continued mentorship and inspiration.

\appendix

    \section{Coherent energy transfer to the cavity}

    We consider an ensemble of two-electronic-state molecular radiators in
    a coherent superposition of their
    ground and excited state at $t_0=0$,
    which is given by Eq.~(\ref{eq:istate}).
    Such collective excitation can
    be achieved by a resonant laser pulser significantly shorter than the
    collective Rabi cycling period of the ensemble.

    The system's Hamiltonian has the general form of
    Eqs.(\ref{eq:htot}-\ref{eq:dipbo1}),
    where an electronic basis
    has been introduced, the second order
    light-matter coupling is left out,
    and the rotating wave approximation is used for the light-matter interaction
    \begin{align}
        \label{eq:hamapp}
        \hat{H} & = \sum_{l=1}^N \Bigg[
            \hat{T}_l \hat{\mathbf{1}}_l +
            |\psi_0^{(l)}\rangle \hat{V}_0 \langle \psi_0^{(l)}| +
            |\psi_1^{(l)}\rangle \hat{V}_1 \langle \psi_1^{(l)}| \\\nonumber
            & + \mu_{01} g
            \Big(
                \hat{a}^\dagger |\psi_0^{(l)}\rangle \langle \psi_1^{(l)}| +
                \hat{a} |\psi_1^{(l)}\rangle \langle \psi_0^{(l)}|
           \Big) \Bigg] \\\nonumber
            & + \hbar\omega_c
            \left(
                \frac{1}{2} + \hat{a}^\dagger\hat{a}
            \right).
    \end{align}
    As discussed in the main text, $\mu_{01}$ is the transition dipole matrix
    element in the electronic basis and $g$ corresponds to the light-matter
    coupling constant in the given cavity.
    For the above Hamiltonian (\ref{eq:hamapp}) and initial state
    (\ref{eq:istate}), with initially zero cavity photons, the
    expectation value of the photon-number in the
    cavity
    \begin{align}
        \label{eq:pnum1}
        \langle N_{ph} \rangle(t) =
        \langle \Psi(0)|
        \hat{U}^\dagger(t) \hat{a}^\dagger \hat{a} \hat{U}(t)
        |\Psi(0)\rangle
    \end{align}
    is calculated by introducing a second-order expansion of the propagator
    \begin{align}
        \label{eq:prop}
        \hat{U}(t) \approx 1 - i\hat{H}t -\frac{1}{2}\hat{H}^2 t^2.
    \end{align}
    After some manipulations this leads to
    \begin{align}
        \label{eq:pnum2}
        \langle N_{ph} \rangle(t)
        & =
        t^2 \mu_{01}^2 g^2
        \bigg(
        \sum_{l=1}^{N}
            p_1 \langle \phi_1^{(l)} | \phi_1^{(l)} \rangle\ \\\nonumber
        & +
        \sum_{\substack{k,l = 1 \\ k\neq l}}^{N}
            \langle \phi_1^{(k)} | \phi_0^{(k)} \rangle
            \langle \phi_0^{(l)} | \phi_1^{(l)} \rangle
            p_0 p_1
        \bigg) \\\nonumber
        & = t^2 \mu_{01}^2 g^2 \bigg(
        N \,p_1 + \left(N^2 -N\right) p_0 p_1 |S_{01}|^2 \bigg),
    \end{align}
    where the last line could be rewritten in terms of the
    off-diagonal entry of the molecular electronic reduced density matrix, using
    \mbox{$p_0 p_1 |S_{01}|^2\equiv |\rho_{01}|^2$} from
    Eq.~(\ref{eq:rho01}).
    In the last line, it is considered that all molecular systems are
    identical and therefore the nuclear wavepacket overlaps
    $S_{01}=\langle \phi_1^{(l)} | \phi_0^{(l)} \rangle$ are idependent of
    molecular index $l$.
    The term in parenthesis in Eq.~(\ref{eq:pnum2}) with nuclear overlap
    $S_{01}=1$ can be readily compared with
    the rate expressions by Dicke in Ref.~\cite{dic54:99}, in
    particular Eq.~(26) for the case $p_1=1$ and with Eq.~(27) for the case
    $p_1=p_0=1/2$.

    The same calculation for initial state
    \begin{align}
        \label{eq:istateP}
        |\Psi(0)\rangle =
        \prod_{l=1}^{N}
        \Bigg(
           & \sqrt{p_0} \phi_0^{(l)}(R_l) |\psi_0^{(l)}\rangle  \\\nonumber
         + & \sqrt{p_1} \phi_1^{(l)}(R_l) |\psi_1^{(l)}\rangle
              \Bigg) |P\rangle,
    \end{align}
    i.e., with $P$ cavity photons already present at $t=0$, generalizes to
    \begin{align}
        \label{eq:pnum3}
        \langle N_{ph} \rangle(t) = P + t^2 \mu_{01}^2 g^2 \bigg(
          & N [(P+1)\,p_1 - P\,p_0] \\\nonumber
        + & \left(N^2 -N\right) p_0 p_1 |S_{01}|^2
          \bigg).
    \end{align}
    An important consequence of this result is that coherent dynamical evolution
    of the molecular ensemble and the cavity mode, which depends on a
    coherent superposition of the molecular electronic states, is completely
    insensitive to the number of photons in the cavity mode.
    Superradiance is as such a spontaneous emission process. On the other hand,
    as expected, the number of cavity
    photons
    enhances absorption and stimulated emission processes.







\end{document}